\documentstyle[twocolumn,eqsecnum,aps,amsmath,epsfig]{revtex}

\renewcommand{\theequation}{\arabic{section}.\arabic{equation}}
\renewcommand{\thefigure}{\arabic{section}.\arabic{figure}}
\renewcommand{\thetable}{\arabic{section}.\arabic{table}}

\begin{document}

\draft
\title{\ \ \ \\
       \vspace{-1.2cm}
       \hfill SAGA-HE-136-98 \\
                           \ \\
                           \ \\
%%\title{Structure functions in the polarized Drell-Yan processes \\
         Structure functions in the polarized Drell-Yan processes \\
         with spin-1/2 and spin-1 hadrons:
         I. general formalism}
\author{S. Hino and S. Kumano \cite{byline}}
\address{Department of Physics, Saga University \\
         Saga, 840-8502, Japan}
%%\date{\today}
\date{Oct. 20, 1998}
\maketitle
\begin{abstract}
We discuss general formalism for the structure functions which can be
investigated in the polarized Drell-Yan processes with
spin-1/2 and spin-1 hadrons. To be specific, the formalism can be
applied to the proton-deuteron Drell-Yan processes.
Because of the spin-1 nature, there are new structure
functions which cannot be studied in the proton-proton reactions.
Imposing Hermiticity, parity conservation, and time-reversal invariance, 
we find that 108 structure functions exist in the Drell-Yan processes.
However, the number reduces to 22 after integrating the cross section
over the virtual-photon transverse momentum $\vec Q_T$ or after taking
the limit $Q_T\rightarrow 0$. There are 11 new structure functions in 
addition to the 11 ones in the Drell-Yan processes of spin-1/2 hadrons.
The additional structure functions are associated with the tensor structure
of the spin-1 hadron, and they could be measured by quadrupole spin
asymmetries. For example, the structure functions exist for ``intermediate"
polarization although their contributions vanish in the longitudinal
and transverse polarization reactions. We show a number of spin
asymmetries for extracting the polarized structure functions.
The proton-deuteron reaction may be realized in the RHIC-Spin project
and other future ones, and it could be a new direction of next
generation high-energy spin physics.
\end{abstract}

\narrowtext

%%%%%%%%%%%%%%%%%%%%%%%%%%%%%%%%%%%%%%%%%%%%%%%%%%%%%%%%%%%%%%%%%%%%%%%%%%%%%%
%%%%%%%%%%%%%%%%%%%%%%%%%%%%%%%%%%%%%%%%%%%%%%%%%%%%%%%%%%%%%%%%%%%%%%%%%%%%%%
\section{Introduction}\label{intro}
\setcounter{equation}{0}

Spin structure of the proton has been investigated through the polarized
deep inelastic lepton scattering. A mysterious aspect of spin physics was
revealed by the European Muon Collaboration (EMC) experimental result 
in 1988 \cite{emcg1}: almost none of the proton spin is carried by quarks. 
However, because the lepton reactions provide us only a limited piece of
information on polarized partons \cite{rhic-j-wg}, we should rely on
other experimental methods such as those of the Relativistic Heavy
Ion Collider (RHIC) -Spin project \cite{rhic-spin}. There, the approved
activities are on various polarized proton-proton (pp) reactions. 

Although the details of the pp reactions should be studied further
\cite{pol-pp}, many pp processes have been investigated for a long time.
On the other hand, there are few studies on the polarized deuteron
reactions in connection with spin-dependent structure functions.
We know that the deuteron has additional spin
structure due to its spin-1 nature. The deuteron target is often
used in the deep inelastic scattering; however, the major purpose is
to extract the ``neutron" structure functions in the deuteron. 
We think that we had better shed light on the deuteron spin structure
itself rather than just using it for finding the neutron information.
Within this context, there are some initial studies on the spin-1
structure functions. In the lepton scattering on the deuteron
\cite{mit-b1,fs}, there exist new tensor structure functions
$b_1$, $b_2$, $b_3$, and $b_4$. Among them, the twist-two structure
functions are $b_1$ and $b_2$, and they are related by the Callan-Gross
type relation $2x b_1=b_2$. 
A phenomenological sum rule was proposed for $b_1$ \cite{ck-b1}
in relation to the electric quadrupole structure.
It could be important also to find the tensor polarization
of sea quarks. Although the sum rule is valid for spin-1 hadrons
within a quark model, there could have complexities in the deuteron
because of nuclear shadowing effects \cite{b1-shadow}.

There are some studies on the polarized lepton-deuteron reaction; however,
the polarized deuteron has not been investigated in hadron-hadron reactions
for finding the polarized structure functions. The general formalism of
the polarized Drell-Yan process of spin-1/2 hadrons was studied
in the pioneering paper of Ralston-Soper \cite{rs} and also the one
by Donoghue and Gottlieb \cite{dg}. In these works, it was revealed
that 48 structure functions can be studied in the reactions of
spin-1/2 hadrons and the number becomes 11 after integrating over
the virtual-photon transverse momentum
$\vec Q_T$ or after taking the limit $Q_T\rightarrow 0$.
The parton-model interpretation of these structure functions
is also discussed in these works \cite{rs,dg} as well as in
the Tangerman-Mulders' paper \cite{tm}. 

The purpose of this paper is to investigate the general formalism
of the polarized Drell-Yan processes with spin-1/2 and spin-1 hadrons
in order to apply it to the polarized proton-deuteron (pd) reactions.
In particular, we discuss what kind of new spin structure functions is
measured and how they are related to the hadron polarizations.
Another purpose is to facilitate future deuteron-spin projects such
as the possible polarized deuteron reactions at RHIC \cite{rhic-d}.

In section \ref{pd-dy}, we discuss general formalism of
the polarized Drell-Yan process with spin-1/2 and spin-1 hadrons
and obtain possible structure functions in the reaction.
Then, we investigate how they are related to the Lorentz index
structure of the hadron tensor in section \ref{hadron}.
Various spin asymmetries are discussed in section \ref{asym}.
 The results are summarized in section \ref{sum}.

\vfill\eject

%%%%%%%%%%%%%%%%%%%%%%%%%%%%%%%%%%%%%%%%%%%%%%%%%%%%%%%%%%%%%%%%%%%%%%%%%%%%%%
%%%%%%%%%%%%%%%%%%%%%%%%%%%%%%%%%%%%%%%%%%%%%%%%%%%%%%%%%%%%%%%%%%%%%%%%%%%%%%
\section{Formalism for the polarized \\
         Drell-Yan process with spin-1/2 \\
         and spin-1 hadrons}
\label{pd-dy}
\setcounter{equation}{0}

In the polarized Drell-Yan processes of spin-1/2 hadrons, possible
structure functions were found by noting the Lorentz index structure
in the hadron tensor \cite{rs}. Then, they were also discussed in a general
framework \cite{dg} by using the Jacob-Wick helicity formalism \cite{jw}.
In order to avoid missing some structure functions in the reactions
of spin-1/2 and spin-1 hadrons,
we first discuss the Donohue-Gottlieb type formalism \cite{dg}
in this section. In section \ref{hadron}, the Ralston-Soper
type formalism is discussed. 

\begin{figure}[h]
   \begin{center}
       \epsfig{file=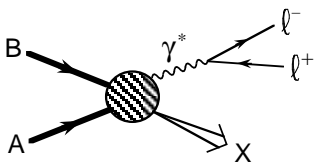,width=4.5cm}
   \end{center}
   \vspace{0.0cm}
   \caption{Drell-Yan process.}
   \label{fig:dy}
\end{figure}

We study the cross section of the Drell-Yan reaction
\begin{equation}
A \, (spin \, 1/2) + B \, (spin \, 1) \rightarrow \ell^+ \ell^- +X
\ ,
\end{equation}
which is schematically shown in Fig. \ref{fig:dy}.
The hadrons A and B are spin-1/2 and
spin-1 particles, respectively. For example, they could be the proton
and the deuteron; however, the following formalism can be applied to
any other hadrons with spin-1/2 and spin-1. 

\vspace{-0.0cm}
%%%%%%%%%%%%%%%%%%%%%%%%%%%%%%%% figure %%%%%%%%%%%%%%%%%%%%%%%%%%%%%%%%%%%%%%
\noindent
%%%%%%%%%%%%%%%%%%%%%%%%%%%%%%%% figure %%%%%%%%%%%%%%%%%%%%%%%%%%%%%%%%%%%%%%

\begin{figure}[h]
   \begin{center}
       \epsfig{file=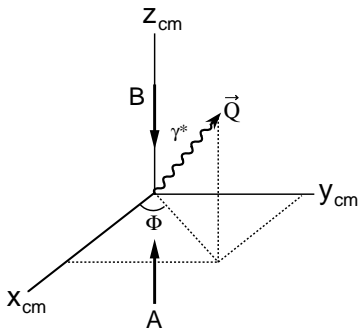,width=5.0cm}
   \end{center}
   \vspace{-0.2cm}
   \caption{Center of momentum frame.}
   \label{fig:cm}
\end{figure}

In describing the cross section, a coordinate system has to be chosen.
The center of momentum (c.m.) frame is easy to be visualized because
it could be the same as or at least close to the laboratory frame
of collider experiments.
At this stage, the polarized deuteron acceleration is not planned
at RHIC \cite{rhic-d}, so that its momentum is not known. 
The c.m. frame is shown in Fig. \ref{fig:cm}. 
\vspace{-0.3cm}
%%%%%%%%%%%%%%%%%%%%%%%%%%%%%%%% figure %%%%%%%%%%%%%%%%%%%%%%%%%%%%%%%%%%%%%%
\noindent
%%%%%%%%%%%%%%%%%%%%%%%%%%%%%%%% figure %%%%%%%%%%%%%%%%%%%%%%%%%%%%%%%%%%%%%%
\noindent
The total dilepton momentum is denoted by $Q$,
and it is expressed in the c.m. frame as
\begin{align}
Q^\mu & \equiv k_{\ell^+}^\mu + k_{\ell^-}^\mu \nonumber \\
      & = (Q_0, Q_T cos \Phi, Q_T sin \Phi, Q_z)_{cm}
\ .
\end{align}
The $z_{cm}$-axis is chosen in the direction of the hadron-A momentum
$\vec P_{A}^{cm}$, so that the hadron-B momentum $\vec P_{B}^{cm}$ 
lies in the $- z_{cm}$ direction. The azimuthal angle 
of $\vec Q$ is denoted as $\Phi$. 

\begin{figure}[h]
   \begin{center}
       \epsfig{file=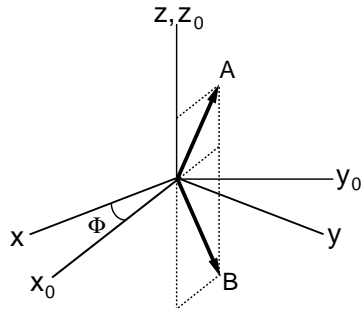,width=5.0cm}
   \end{center}
   \vspace{-0.2cm}
   \caption{Dilepton rest frame.}
   \label{fig:rest}
\end{figure}

The theoretical description becomes simple if the dilepton rest frame,
rather than the c.m. frame, is chosen. It literally means that the
total dilepton momentum vanishes ($\vec Q=0$) in the frame. 
In order to obtain this frame from the c.m. in Fig. \ref{fig:cm},
the frame has to be boosted first to the $z_{cm}$
direction so as to get $Q_z=0$, then to the $\vec Q_T$
direction so as to obtain $\vec Q=0$. The resulting frame is
shown by $x$, $y$, and $z$ axes in Fig. \ref{fig:rest}.
The momenta $\vec P_{A}$ and $\vec P_{B}$ are no longer along
the $z$ axis. In the following formalism of this section, we take
the dilepton rest frame as the Collins-Soper frame \cite{cs},
which is shown by $x_0$, $y_0$, and $z_0$ axes in Fig. \ref{fig:rest}.
\vspace{-0.3cm}
%%%%%%%%%%%%%%%%%%%%%%%%%%%%%%%% figure %%%%%%%%%%%%%%%%%%%%%%%%%%%%%%%%%%%%%%
\noindent
%%%%%%%%%%%%%%%%%%%%%%%%%%%%%%%% figure %%%%%%%%%%%%%%%%%%%%%%%%%%%%%%%%%%%%%%
\noindent
The $z_0$ axis is chosen so as to bisect the angle between
the momenta $\vec P_{A}$ and $-\vec P_{B}$:
\begin{equation}
\hat z_0 = \frac{\vec P_{A} (P_{B} \cdot Q) - 
               \vec P_{B} (P_{A} \cdot Q)}
              {| \vec P_{A} (P_{B} \cdot Q) - 
                 \vec P_{B} (P_{A} \cdot Q) |}
\ .
\end{equation}
Throughout this paper, the $g_{\mu\nu}$ convention
$g_{00}=-g_{11}=-g_{22}=-g_{33}=1$ is used. 
Then the $y_0$-axis is taken as
\begin{equation}
\hat y_0 = - \frac{  \vec P_{A} \times \vec P_{B}  }
                {| \vec P_{A} \times \vec P_{B} |}  
\ .
\end{equation}
The remaining $x_0$-axis is chosen so that it is orthogonal 
to $\hat y_0$ and $\hat z_0$. From the above definitions of
$\hat y_0$ and $\hat z_0$, $\hat x_0$ is equal to the transverse
unit vector $\hat Q_T$ which is in the direction away
from $(\vec P_{A}+ \vec P_{B})_{_T}$.
Between the two dilepton rest frames,
there is a difference of the azimuthal angle $\Phi$
as shown in Fig. \ref{fig:rest}. This difference becomes
significant when the cross section is integrated over $\vec Q_T$ later.

The Drell-Yan cross section is given by
\begin{multline}
d \sigma = \frac{1}{4 \sqrt{(P_{A} \cdot P_{B})^2 -M_A^2 M_B^2}}
       \sum_{s_{\ell^+}, s_{\ell^-}} \sum_X 
\\
    \times (2\pi)^4 
           \delta^4 (P_{A}+P_{B}-k_{\ell^+}-k_{\ell^-}-P_{X})
\\
    \times \ | < \ell^+ \ell^- X | T | A B > |^2
               \frac{d^3 k_{\ell^+}}{(2\pi)^3 2 E_{\ell^+}}  \,
               \frac{d^3 k_{\ell^-}}{(2\pi)^3 2 E_{\ell^-}}
\ ,
\end{multline}
where $M_A$ and $M_B$ are A and B masses, 
$s_{\ell^+}$ and $s_{\ell^-}$ indicate the $\ell^+$ and $\ell^-$ spins,
$s$ is the center-of-mass energy squared $s=(P_A+P_B)^2$,
and the matrix element is given by
\begin{multline}
< \ell^+ \ell^- X | T | A B >
 \, = \, \bar u(k_{\ell^-},s_{\ell^-}) e\gamma_\mu 
              v(k_{\ell^+},s_{\ell^+}) \, 
\\ \times
                     \frac{g^{\mu\nu}}{(k_{\ell^+}+k_{\ell^-})^2}  \,
                     < X | e J_\nu(0) | A B >
\ .
\end{multline}
In the dilepton rest frame, the cross section becomes
\begin{equation}
\frac{d\sigma}{d^4 Q d \Omega} = \frac{\alpha^2}{2 \, s \, Q^4} 
                                    \, L_{ij} \, W_{ij}
\ ,
\label{eqn:cross0}
\end{equation}
where $\alpha=e^2/(4\pi)$ is the fine structure constant,
the component $L_{00}$ vanishes in this frame, 
and the hadron and lepton masses are
neglected by considering $M_A^2, M_B^2 \ll s$ and $m_\ell^2 \ll Q^2$.
The lepton part is given by
\begin{align}
L_{ij} & = \frac{1}{2} \sum_{s_{\ell^+}, s_{\ell^-}} 
                     [ \, u(k_{\ell^-},s_{\ell^-}) \gamma_i 
                          v(k_{\ell^+},s_{\ell^+}) \, ]^\dagger
\nonumber \\ 
      & \ \ \ \ \ \ \ \ \ \  \ \ \ \times 
                     [ \, u(k_{\ell^-},s_{\ell^-}) \gamma_j 
                          v(k_{\ell^+},s_{\ell^+}) \, ]
\nonumber \\
      & = 4 \, \vec k^{\, 2} \left ( \delta_{ij} - 
                              k_i k_j /|\vec k|^2 \right )
\ ,
\end{align}
where $\vec k\equiv \vec k_{\ell^+}$ and the lepton mass
terms are neglected ($m_\ell^2 \ll \vec k^{\, 2}$).
The hadron part is given by
\begin{equation}
W_{ij} = \int \frac{d^4\xi}{(2\pi)^4} \, e^{iQ \cdot \xi}
            < \! AB \, | \, J_i (0) \, J_j (\xi) \, | \, AB \! >
\ .
\end{equation}
The polar and azimuthal angles $\theta$ and $\phi_0$
of the lepton $\ell^+$ are defined by
\begin{equation}
\hat k \cdot \hat z_0 = cos \theta, \ \ \ 
\hat k \cdot \hat y_0 = sin \theta \, sin \phi_0
\ ,
\end{equation}
where $\hat k= \vec k/ |\vec k|$. It is convenient to express
the lepton tensor $L_{ij}$ as
\begin{multline}
L_{ij} = \frac{8 \, \vec k^{\, 2}}{3} 
         \sum_{\lambda,\lambda',L,M} f(L) 
          < \! 1 \lambda : LM | 1 \lambda' \! >
\\ \times
          D_{M0}^{\ L} (\phi_0,\theta,0) \, 
           \varepsilon_i (\lambda') \, \varepsilon_j^* (\lambda)
\ ,
\end{multline}
where $< 1 \lambda : LM | 1 \lambda'>$ is a Clebsch-Gordan coefficient
and $\varepsilon_i (\lambda)$ is the $i$ component of
the spherical unit vector \cite{edmonds}.
The coefficients $f(L)$ are defined by
\begin{equation}
f(0)=1 \, , \ \ \  f(1)=0 \, , \ \ \ f(2)=\frac{\sqrt{10}}{2}
\, .
\end{equation}
If the lepton mass cannot be neglected, $f(2)$ becomes
the expression in Ref. \cite{dg}.
We use the rotational matrix element $D_{m m'}^{\ j} (\alpha,\beta,\gamma)$
which is defined by \cite{fesh}
\begin{align}
D_{m m'}^{\ j} (\alpha,\beta,\gamma)
         & \equiv \, < jm \, | \, e^{-i\alpha J_z} \,
             e^{-i\beta J_y} \, e^{-i\gamma J_z} \, | \, jm'>
\nonumber \\
         & = e^{-im\alpha} \, d_{m m'}^{\ j} (\beta) \, e^{-im'\gamma}
\ .
\label{eqn:d-def}
\end{align}
We should be careful about the definition of the $D$-matrix
because there are different conventions.
The definition of Eq. (\ref{eqn:d-def}) is used throughout
this paper; however, it is different, for example,
from the one in Ref. \cite{edmonds}.

Using these expressions and introducing spin density matrix, 
the cross section becomes
\begin{multline}
\frac{d\sigma}{d^4 Q d \Omega_0} = \frac{1}{4 \, \pi} 
      \sum_{L,M} f(L) D_{M0}^{\ L} (\phi_0,\theta,0)  \, 
     \sum_{\lambda,\lambda'} \sum_X 
\\ \times
     \delta^4 (P_{A}+P_{B}-Q-P_{X}) \, 
     < 1 \lambda : LM | 1 \lambda'> 
\\ \times
     \sum_{\mu,\mu'} \sum_{\nu,\nu'} \,
     \rho (AB)_{\mu\nu:\mu ' \nu '} \,
     F_{\lambda ' \mu ' \nu '}^\dagger  \, 
     F_{\lambda\mu\nu}
\label{eqn:cross1}
\ .
\end{multline}
The helicity amplitude $F_{\lambda\mu\nu}$ is given by
\begin{equation}
F_{\lambda\mu\nu} = \sqrt{\frac{\pi \, \alpha^2}
                               {3 \, s \, \vec k^{\, 2}}} \, 
         <X \, | \, \vec \varepsilon^{\, \, *} (\lambda) \cdot
                   \vec J (0) \, | \, A(\mu) B(\nu) >
\ ,
\end{equation}
and the spin density matrix $\rho (AB)_{\mu\nu:\mu ' \nu '}$ is
\begin{equation}
\rho (AB)_{\mu\nu:\mu ' \nu '} = 
    \rho (A)_{\mu\mu '} \, \rho (B)_{\nu\nu '}
\ .
\label{eqn:rhoab}
\end{equation}
The density matrix for the hadron $A$ or $B$ is defined by
\begin{equation}
\rho = \frac{1}{2S+1} \sum_{K,N} \, <\tau_{_{KN}} (S) > \, 
         \tau_{_{KN}}^\dagger (S)
\ , 
\end{equation}
where the statistical tensor $\tau_{_{KN}} (S)$ is expressed as
\cite{kk-book}
\begin{multline}
\tau_{_{KN}} (S) = \sqrt{2S+1} \sum_{\mu,\mu'} \, (-1)^{S-\mu} \,
                < S \mu' : S -\mu | KN >
\\ \times
             |S\mu'> <S\mu| 
\label{eqn:tau}
\ .
\end{multline}
Therefore, the density matrix becomes
\begin{align}
\rho_{\mu\mu'} = & < S \mu \, | \, \rho \, | \, S \mu' > 
\nonumber \\
      = & \sum_{K,N} \, \frac{\sqrt{2K+1}}{2S+1} \,
         < S \mu' : KN | S \mu > \, 
         < \tau_{_{KN}}^\dagger (S) >
\ .
\label{eqn:rho-tau}
\end{align}
The quantities $< \! \tau_{_{KN}}^\dagger (S) \! >$ 
for the spin 1/2 and 1 particles are
\begin{align}
< \tau_{_{00}}^\dagger (S) >
      & =  \, 1
\ , 
\\
< \tau_{_{1N}}^\dagger (1/2) >
      & =  \, | \vec S | \, D_{N 0}^{\ 1} (\alpha,\beta,0)
\ ,
\label{eqn:tau1}
\\
< \tau_{_{1N}}^\dagger (1) > 
      & = \sqrt{\frac{3}{2}} \, | \vec S | \, 
                        D_{N 0}^{\ 1} (\alpha,\beta,0)
\ ,
\label{eqn:tau2}
\end{align}
where $\alpha$ and $\beta$ are azimuthal and polar angles of
the polarization vector $\vec S$. 
In addition to these tensors, there exists another one for the
spin-1 particle. As it is obvious from the definition
of Eq. (\ref{eqn:tau}), the rank-two ($K=2$) tensor is possible.
We also express it in terms of the $D$-matrix:
\begin{equation}
<\tau_{2N}^\dagger (1) >  
      \, = \, 
    < \sqrt{2} \, \vec S^{\, 2} \, D_{N 0}^{\ 2} (\alpha,\beta,0) >
\ .
\label{eqn:tau3}
\end{equation}
%%%%%%%%%%%%%%%%%%%%%%%%%%%%%%%%%%%%%%%%%
\vfill\eject\widetext\noindent
%%%%%%%%%%%%%%%%%%%%%%%%%%%%%%%%%%%%%%%%%
However, special attention should be paid for $<\vec S^{\, 2}>=2$
in calculating 
$<\tau_{20}^\dagger (1) > 
     = < \vec S^{\, 2} \, (3\, cos^2 \beta - 1) / \sqrt{2} >
     = $ 
$< ( 3 \, S_z^2 - \vec S^{\, 2} ) / \sqrt{2} >$.
Furthermore, $<S_i \, S_j >$ should be replaced by
$< S_i \, S_j + S_j \, S_i >/2$.
For simplicity, the notation $<>$ of Eq. (\ref{eqn:tau3})
is not written in the following equations. 
The polarization angles are denoted as $\alpha_{_A}$ and $\beta_{_A}$
for the hadron $A$ and as $\alpha_{_B}$ and $\beta_{_B}$ for $B$.
The angles $\alpha_{_B}$ and $\beta_{_B}$ ($\alpha_{_A}$ and $\beta_{_A}$)
are measured by taking the coordinates as 
$x_{_B}=x_{cm}$, $y_{_B}=-y_{cm}$, and $z_{_B}=-z_{cm}$ 
($x_{_A}=x_{cm}$, $y_{_A}=y_{cm}$, and $z_{_A}=z_{cm}$). 
As it is pointed out in Ref. \cite{dg}, the transverse axis
could be chosen arbitrary. The azimuthal angles
$\Phi$, $\alpha_{_A}$, and $\alpha_{_B}$ transform into
$\Phi-\varepsilon$, $\alpha_{_A}-\varepsilon$, 
and $\alpha_{_B}+\varepsilon$ under the angle-$\varepsilon$ rotation
about the $z$ axis, so that only the angles $\alpha_{_A}-\Phi$
and $\alpha_{_B}+\Phi$ have physical meaning. 
In this way, the angles $\alpha_{_A}$ and $\alpha_{_B}$ are
replaced by $\alpha_{_A}-\Phi$ and $\alpha_{_B}+\Phi$, respectively
in the expressions of the $D$-matrices.       
It is noteworthy to emphasize that new $\vec S^{\, 2}$ terms exist
in our Drell-Yan process whereas merely the linear terms
of $\vec S$ are enough for the pp reaction. 
Therefore, the study of the new spin structure is 
in other words to investigate the rank-two $\vec S^{\, 2}$ terms. 

Substituting these expressions of $< \! \tau_{_{KN}}^\dagger (S) \! >$ 
into Eq. (\ref{eqn:rho-tau}), 
we obtain the spin density matrix for the reaction
in Eq. (\ref{eqn:rhoab}).
Then, the cross section of Eq. (\ref{eqn:cross1}) becomes
\begin{align}
\frac{d\sigma}{d^4 Q d \Omega_0} = 
                     \frac{1}{4\pi} \sum_{L = 0, 2} & f(L) 
                     \sum_{M = -L}^L
                     D_{M 0}^{\ L} (\phi_0, \theta, 0) \,
                     \biggr[ \, R_L^M {}_0^0 {}_0^0 
                     + \sqrt3 \, |\vec{S}_A| \sum_{N_1} 
                               R_L^M {}_{\ 1}^{N_1} {}_0^0   \, 
                          D_{N_1 0}^{\ 1} (\alpha_A - \Phi, \beta_A, 0)
\nonumber \\ 
                    &+ \frac{3}{\sqrt 2} \, |\vec{S}_B| \sum_{N_2}
                               R_L^M {}_0^0 {}_{\ 1}^{N_2} \, 
                            D_{N_2 0}^{\ 1} (\alpha_B + \Phi, \beta_B, 0) 
                     + \sqrt{10}  \, \vec{S}_B^{\, 2}
                               \sum_{N_2} R_L^M {}_0^0 {}_{\ 2}^{N_2} \, 
                            D_{N_2 0}^{\ 2} (\alpha_B + \Phi, \beta_B, 0)
\nonumber \\ 
                 &+ \frac{3 \sqrt{3}}{\sqrt{2}} \, |\vec{S}_A| \, |\vec{S}_B|
                               \sum_{N_1, N_2} 
                               R_L^M {}_{\ 1}^{N_1} {}_{\ 1}^{N_2} \, 
                            D_{N_1 0}^{\ 1} (\alpha_A - \Phi, \beta_A, 0) \,
                            D_{N_2 0}^{\ 1} (\alpha_B + \Phi, \beta_B, 0) 
\nonumber \\ 
                    &+ \sqrt{30} \, |\vec{S}_A| \, \vec{S}_B^{\, 2}
                               \sum_{N_1, N_2} 
                               R_L^M {}_{\ 1}^{N_1} {}_{\ 2}^{N_2}  \, 
                           D_{N_1 0}^{\ 1} (\alpha_A - \Phi, \beta_A, 0) \,
                           D_{N_2 0}^{\ 2} (\alpha_B + \Phi, \beta_B, 0) \,
                     \biggr]
\ ,
\label{eqn:cross2}
\end{align}
where the structure function $R_{L K_1 K_2}^{M N_1 N_2}$ is defined by
the helicity amplitudes and the Clebsch-Gordan coefficients as
\begin{align}
R_L^M {}_{K_1}^{N_1} {}_{K_2}^{N_2} \equiv \, \frac{1}{6} \,
                      \sum_{\lambda , \mu , \nu} \sum_X \, & \,
                      \delta^4 (P_A + P_B - Q - P_X) \,
                       < 1 \lambda : LM | 1 \lambda'> 
\nonumber \\ \times \, 
 &               F_{\lambda' \mu' \nu'}^\dagger  \, F_{\lambda \mu \nu}
                     <\frac{1}{2} \mu': K_1 N_1| \frac{1}{2} \mu> \, 
                     <1 \nu': K_2 N_2| 1 \nu>
\ .                                          
\end{align}
The terms which do not exist in the pp reactions are those proportional to 
$\vec{S}_B^{\, 2}$ and $|\vec{S}_A| \, \vec{S}_B^{\, 2}$ 
in Eq. (\ref{eqn:cross2}). These terms are, therefore, associated with
the new spin structure for the spin-1 hadron. 
According to Eq. (\ref{eqn:cross2}), the structure functions 
$R_L^M {}_0^0 {}_{\ 2}^{N_2}$ and $R_L^M {}_{\ 1}^{N_1} {}_{\ 2}^{N_2}$
appear as the additional ones.

We have shown that new structure functions exist in the polarized
Drell-Yan reactions with spin-1/2 and spin-1 hadrons. 
However, it is not still clear how many
structure functions could be studied in the reactions because
the cross section is written in the summation form.
Many of the possible functions $R_L^M {}_{K_1}^{N_1} {}_{K_2}^{N_2}$
are not independent, and the number can be reduced by
imposing Hermiticity, parity conservation, and time-reversal invariance.
The Hermiticity requires 
\begin{equation}
R_L^M {}_{K_1}^{N_1} {}_{K_2}^{N_2} 
     = (-1)^{M + N_1 + N_2} 
       \left( R_{\ L}^{-M} {}_{\ K_1}^{-N_1} {}_{\ K_2}^{-N_2} 
       \right)^*
\ .
\end{equation}
The parity conservation and time-reversal invariance require
\begin{equation}
R_L^M {}_{K_1}^{N_1} {}_{K_2}^{N_2} 
     = (-1)^{L + K_1 + K_2 + M + N_1 + N_2} 
       \, R_{\ L}^{-M} {}_{\ K_1}^{-N_1} {}_{\ K_2}^{-N_2} 
\ .
\end{equation}
These equations indicate
$ \left( R_{L}^{M} {}_{K_1}^{N_1} {}_{K_2}^{N_2} \right)^*
 =(-1)^{L + K_1 + K_2} \, R_{L}^{M} {}_{K_1}^{N_1} {}_{K_2}^{N_2}$.
Therefore, the structure functions have the property
\begin{equation}
R_L^M {}_{K_1}^{N_1} {}_{K_2}^{N_2}  = 
  \begin{cases}
    \text{real}      &  \text{if $L + K_1 + K_2$=even number,} \\
    \text{imaginary $\ \ $} & \text{if $L + K_1 + K_2$=odd number,} \\
    \text{ 0 }       &  \text{if $L + K_1 + K_2$=odd number and
                                $M + N_1 + N_2 = 0$.}
  \end{cases} 
\end{equation}   
Imposing these conditions on the structure functions
$R_L^M {}_{K_1}^{N_1} {}_{K_2}^{N_2}$,
we find that 108 independent structure functions exist
in our Drell-Yan processes. 
In the reactions of spin-1/2 hadrons, 
there are 48 structure functions. 
Therefore, there exist 60 additional ones in the case of
spin-1/2 and spin-1 hadrons.
%%%%%%%%%%%%%%%%%%%%%%%%%%%%%%%%%%%%%%%%%
\vfill\eject

$\ \ \ $

\vfill\eject
%%%%%%%%%%%%%%%%%%%%%%%%%%%%%%%%%%%%%%%%%
The polarized pp Drell-Yan processes will be investigated
at RHIC; however, all of the 48 structure functions 
for the spin-1/2 hadrons would not be measured.
Of course, all of them are not important at this stage.
We know that even the leading-twist contributions are not well
understood yet. The essential structure functions in our reactions
can be extracted by integrating over the transverse momentum
$\vec Q_T$. However, the integration over the azimuthal angle $\Phi$
of $\vec Q_T$ is sufficient in order
to be compared with the result in section \ref{hadron}.
We have to be careful about the coordinates in calculating the integration
because our present coordinate system is taken so that the $x_0$-axis
is in the direction of $\vec Q_T$. The $x_0$ and $y_0$ axes
should be rotated around $z$ when the integral of $\Phi$ is calculated.
We fix the frame as the $(x,y,z)$ coordinates
in Fig. \ref{fig:rest}, then the difference between the azimuthal
angles is $\Phi$ as shown in the figure:
\begin{equation}
\frac{d\sigma}{d^4 Q \, d \Omega} =
\left.
 \frac{d\sigma}{d^4 Q \, d \Omega_0} \right | _{\phi_0 = \phi - \Phi}
\ .
\end{equation}
After the $\Phi$ integration, the only terms with $M=N_2-N_1$
remain and the cross section becomes
\begin{align}
2 \,  \int d\Phi \, \frac{d\sigma}{d^4 Q \, d \Omega} =&
          \sum_{L=0,2} f(L) \sum_{M = -L}^L 
                  D_{M 0}^{\ L} (\phi, \theta, 0) 
  \,  \biggr[ \, \delta_{M, 0} \, R_L^M {}_0^0 {}_0^0 
         + \sqrt{3} \, |\vec{S}_A| \, R_L^M {}_{\ \ 1}^{-M} {}_0^0 
              \,  D_{-M 0}^{\ 1} (\alpha_A, \beta_A, 0) 
\nonumber \\ 
         &+ \frac{3}{\sqrt{2}} \, |\vec{S}_B| \,  R_L^M {}_0^0 {}_{\ 1}^M  
                \,  D_{M 0}^{\ 1} (\alpha_B, \beta_B, 0)  
        + \sqrt{10} \, \vec{S}_B^{\, 2} \, R_L^M {}_0^0 {}_{\ 2}^M 
                \,  D_{M 0}^{\ 2} (\alpha_B, \beta_B, 0) 
\nonumber \\ 
         &+ \frac{3\sqrt{3}}{\sqrt{2}} \, |\vec{S}_A| \, |\vec{S}_B|
                  \sum_{N_1, N_2} \delta_{M, N_2 - N_1} \, 
                  R_L^M {}_{\ 1}^{N_1} {}_{\ 1}^{N_2} \, 
                  D_{N_1 0}^{\ 1} (\alpha_A, \beta_A, 0)  \,
                  D_{N_2 0}^{\ 1} (\alpha_B, \beta_B, 0) 
\nonumber \\ 
         &+ \sqrt{30} \, |\vec{S}_A| \, \vec{S}_B^{\, 2}
                  \sum_{N_1, N_2} \delta_{M, N_2 - N_1}  \, 
                  R_L^M {}_{\ 1}^{N_1} {}_{\ 2}^{N_2}    \,
                  D_{N_1 0}^{\ 1} (\alpha_A, \beta_A, 0)     \, 
                  D_{N_2 0}^{\ 2} (\alpha_B, \beta_B, 0)   \, 
          \biggr]     
\ .
\label{eqn:cross3}      
\end{align}               
There exist 22 structure functions in this equation, and
they are physically significant ones which could be investigated
experimentally. In order to clarify the situation, we list
these 22 structure functions in Table \ref{tab:list}.

\narrowtext

\begin{table}
\caption{List of the possible structure functions 
         $R_L^M {}_{K_1}^{N_1} {}_{K_2}^{N_2}$ after 
         the integration over $\Phi$.
         The asterisk * indicates a new structure function
         which does not exist in the Drell-Yan processes
         of spin-1/2 hadrons.
   \label{tab:list}}
\begin{tabular}{|ccccccc|}
    $R_L^M {}_{K_1}^{N_1} {}_{K_2}^{N_2}$
            & $L$ & $M$ & $K_1$ & $N_1$ & $K_2$ & $N_2$   \\
\hline
           &  0  &  0  &   0   &   0   &   0   &   0     \\   
         * &  0  &  0  &   0   &   0   &   2   &   0     \\   
           &  0  &  0  &   1   &   0   &   1   &   0     \\   
           &  0  &  0  &   1   &   1   &   1   &   1     \\   
         * &  0  &  0  &   1   &   1   &   2   &   1     \\   
           &  2  &  0  &   0   &   0   &   0   &   0     \\   
         * &  2  &  0  &   0   &   0   &   2   &   0     \\   
           &  2  &  0  &   1   &   0   &   1   &   0     \\   
           &  2  &  0  &   1   &   1   &   1   &   1     \\   
         * &  2  &  0  &   1   &   1   &   2   &   1     \\   
           &  2  &  1  &   0   &   0   &   1   &   1     \\   
         * &  2  &  1  &   0   &   0   &   2   &   1     \\   
           &  2  &  1  &   1   & $-1$  &   0   &   0     \\   
           &  2  &  1  &   1   & $-1$  &   1   &   0     \\   
         * &  2  &  1  &   1   & $-1$  &   2   &   0     \\   
           &  2  &  1  &   1   &   0   &   1   &   1     \\   
         * &  2  &  1  &   1   &   0   &   2   &   1     \\   
         * &  2  &  1  &   1   &   1   &   2   &   2     \\   
         * &  2  &  2  &   0   &   0   &   2   &   2     \\   
           &  2  &  2  &   1   & $-1$  &   1   &   1     \\   
         * &  2  &  2  &   1   & $-1$  &   2   &   1     \\   
         * &  2  &  2  &   1   &   0   &   2   &   2     \\  
\end{tabular}
\end{table}
%%%%%%%%%%%%%%%%%%%%%%%%%%%%%%%%%%%%%%%%%
\vfill\eject

$\ \ \ $

\vfill\eject
%%%%%%%%%%%%%%%%%%%%%%%%%%%%%%%%%%%%%%%%%

\widetext\noindent
The 11 structure functions with the asterisk mark *
in Table \ref{tab:list} are new ones which are
associated with the spin-1 nature of the hadron $B$.
The other 11 structure functions exist in the
Drell-Yan processes of spin-1/2 hadrons, so that the interesting
point of our reactions (e.g. the polarized proton-deuteron processes)
is to investigate the details of the new 11 structure functions.

Substituting explicit expressions for the $D$-matrices
and calculating the the summations over $L$, $M$, $N_1$, and $N_2$,
we have the cross section with the 22 structure functions:
\begin{align}
2 \, \int d\Phi \, & \frac{d\sigma}{d^4 Q \, d \Omega}  = 
      f(0) \, \biggr[ \, R_0^0 {}_0^0 {}_0^0
             +\frac{3 \sqrt{3}}{\sqrt{2}} \, |\vec{S}_A| \, |\vec{S}_B| \, 
               \bigg\{ \, cos \beta_A \, cos \beta_B \, R_0^0 {}_1^0 {}_1^0
                 +sin \beta_A \, sin \beta_B \, cos(\alpha_A+\alpha_B) \,
                        R_0^0 {}_1^1 {}_1^1 \, \bigg\}
\nonumber \\ & \ \ \ \ \ \ \ \ \ \ \ 
               +\frac{\sqrt{5}}{\sqrt{2}} \, \vec{S}_B^{\, 2} \,
                  (3\, cos^2 \beta_B-1) \, R_0^0 {}_0^0 {}_2^0
               - 3 \sqrt{10} \, |\vec{S}_A| \, \vec{S}_B^{\, 2} \, 
                    sin \beta_A \, sin \beta_B \, cos \beta_B \, 
                    sin (\alpha_A+\alpha_B) \, i \, R_0^0 {}_1^1 {}_2^1 
                \, \biggr]
\nonumber \\ &
     +f(2) \, (3\, cos^2 \theta-1) \, \biggr[ \, \frac{1}{2} \, 
                            R_2^0 {}_0^0 {}_0^0
             +\frac{3 \sqrt{3}}{2 \sqrt{2}} \, |\vec{S}_A| \, |\vec{S}_B| \, 
               \bigg\{ \, cos \beta_A \, cos \beta_B \, R_2^0 {}_1^0 {}_1^0
                 +sin \beta_A \, sin \beta_B \, cos(\alpha_A+\alpha_B) \,
                        R_2^0 {}_1^1 {}_1^1 \, \bigg\}
\nonumber \\ & \ \ \ \ \ \ \ \ \ \ \ 
               +\frac{\sqrt{5}}{2 \sqrt{2}} \, \vec{S}_B^{\, 2} \, 
                   (3\, cos^2 \beta_B-1) \,R_2^0 {}_0^0 {}_2^0
               -\frac{3 \sqrt{5}}{\sqrt{2}} \, |\vec{S}_A| \,
                    \vec{S}_B^{\, 2} \, 
                    sin \beta_A \, sin \beta_B \, cos \beta_B \, 
                    sin (\alpha_A+\alpha_B) \, i \, R_2^0 {}_1^1 {}_2^1 
                \, \biggr]
\nonumber \\ &
     +f(2) \, sin \theta \, cos \theta \, \biggr[ \, 
             sin (\phi-\alpha_A) \, 
               \bigg\{ 3  \, |\vec{S}_A| \, sin \beta_A \, 
                          i \, R_2^1 {}_{\ 1}^{-1} {}_0^0
               +\frac{3 \sqrt{5}}{\sqrt{2}} \, |\vec{S}_A| \, \vec{S}_B^2 \, 
                         sin \beta_A \, (3\, cos^2 \beta_A-1) \, 
                         i \, R_2^1 {}_{\ 1}^{-1} {}_2^0
               \bigg\}
\nonumber \\ & \ \ \ \ \ \ \ \ \ \ \ \ \ \ \ \ \ \ \ \ \ \ 
           - sin (\phi+\alpha_B) \, 
           \bigg\{ \frac{3 \sqrt{3}}{\sqrt{2}} \, |\vec{S}_B| \, sin \beta_B \, 
                        i \, R_2^1 {}_0^{0} {}_1^1
             + 3 \sqrt{30} \, |\vec{S}_A| \, \vec{S}_B^{\, 2} \, 
                     cos \beta_A \,  sin \beta_B \,  cos \beta_B \, 
                         i \, R_2^1 {}_1^{0} {}_2^1
               \bigg\}
\nonumber \\ & \ \ \ \ \ \ \ \ \ \ \ \ \ \ \ \ \ \ \ \ \ \  
          - sin (\phi + \alpha_A + 2 \alpha_B ) \, \frac{3 \sqrt{15}}{2} \, 
              |\vec{S}_A| \, \vec{S}_B^{\, 2} \, sin \beta_A \, sin^2 \beta_B \, 
                  i \, R_2^1 {}_1^1 {}_2^2
\nonumber \\ & \ \ \ \ \ \ \ \ \ \ \ \ \ \ \ \ \ \ \ \ \ \ 
          - cos (\phi - \alpha_A ) \, \frac{9}{\sqrt{2}} \, 
                  |\vec{S}_A| \, |\vec{S}_B| \, sin \beta_A \, cos \beta_B \, 
                   R_2^1 {}_{\ 1}^{-1} {}_1^0
\nonumber \\ & \ \ \ \ \ \ \ \ \ \ \ \ \ \ \ \ \ \ \ \ \ \ 
           + cos (\phi+\alpha_B) \, 
       \bigg\{ 3 \sqrt{10} \, \vec{S}_B^{\, 2} \, 
                   sin \beta_B \, cos \beta_B \, 
                           R_2^1 {}_0^{0} {}_2^1
               +\frac{9}{\sqrt{2}} \, |\vec{S}_A| \, |\vec{S}_B| \, 
                         cos \beta_A \,  sin \beta_B \, 
                         \, R_2^1 {}_1^{0} {}_1^1
               \bigg\} \biggr]
\nonumber \\ &
     +f(2) \, sin^2 \theta \, \biggr[ \, 
             cos (2 \phi + 2 \alpha_B) \, 
             \frac{3 \sqrt{5}}{2 \sqrt{2}} \, 
                  \vec{S}_B^{\, 2} \, sin^2 \beta_B \,
                         R_2^2 {}_0^{0} {}_2^2
           - cos (2 \phi -\alpha_A + \alpha_B) \, 
             \frac{9}{4} \, |\vec{S}_A| \, |\vec{S}_B| \, 
             sin \beta_A \, sin \beta_B \,  
                         R_2^2 {}_{\ 1}^{-1} {}_1^1
\nonumber \\ & \ \ \ \ \ \ \ \ \ \ \ \ \ \ \ \ \ \ \ \ \ \ 
          + sin (2 \phi - \alpha_A + \alpha_B ) \, \frac{3 \sqrt{15}}{2} \, 
          |\vec{S}_A| \, \vec{S}_B^{\, 2} \, sin \beta_A \, 
          sin \beta_B \, cos \beta_B \,
                  i \, R_2^2 {}_{\ 1}^{-1} {}_2^1
\nonumber \\ & \ \ \ \ \ \ \ \ \ \ \ \ \ \ \ \ \ \ \ \ \ \ 
          - sin (2 \phi + 2 \alpha_B ) \, \frac{3 \sqrt{15}}{2\sqrt{2}} \, 
              |\vec{S}_A| \, \vec{S}_B^{\, 2} \, cos \beta_A \, sin^2 \beta_B \,
                  i \, R_2^2 {}_1^{0} {}_2^2
            \, \bigg ]
\ .
\label{eqn:cross4}      
\end{align} 
Here, the term $\vec S_B^{\, 2} (3\, cos\beta_B^2-1)$
should be replaced by 
$3\, <S_{Bz}^2> - <\vec S_B^{\, 2}> = 3\, <S_{Bz}^2> - 2$ as it was explained
after Eq. (\ref{eqn:tau3}).
It is interesting to find that the new structure functions with $K_2=2$
are multiplied by the polarization-angle factors,
$3\, cos^2 \beta_B-1$, $sin^2 \beta_B$, and $sin \beta_B \, cos \beta_B$.
Because the factor $3\, cos^2 \beta_B-1 \sim Y_{20}$ is 
usually associated with the quadrupole structure of the spin-1
hadrons, it is easy to see that these structure functions are related
to such tensor structure. In the same way, other factors are also related
to the spherical harmonics as $sin^2 \beta_B \sim Y_{22}$ and
$sin \beta_B \, cos \beta_B \sim Y_{21}$, so that these structure functions
are also related to the tensor structure. It is particularly interesting
to find the factor $sin\beta_B \, cos\beta_B = sin (2 \beta_B )/2$ 
in the cross section. It means that the structure functions with
$sin \beta_B \, cos \beta_B$ {\it cannot} be found in the longitudinal
($\beta_B=0$) or transverse ($\beta_B=\pi/2$) polarization experiments.
In order to measure them, we should use the polarization
condition $0 < \beta_B < \pi/2$ or $\pi/2 < \beta_B < \pi$. For example,
$\beta_B = \pi/4$ is an appropriate choice. We call this polarization
{\it intermediate polarization} in the sense that it is between
the longitudinal and transverse polarization states.
The longitudinal, transverse, and intermediate polarization
states are denoted as $L$, $T$, and $I$, respectively
in the following discussions.

In this section, we have shown that 108 structure functions exist
in the Drell-Yan processes with spin-1/2 and spin-1 hadrons.
After the integration over the azimuthal angle $\Phi$ (or over $\vec Q_T$),
22 structure functions can be investigated.
Among them, there are 11 new structure functions which do not exist
in the reactions of spin-1/2 hadrons.

%%%%%%%%%%%%%%%%%%%%%%%%%%%%%%%%%%%%%%%%%
\vfill\eject

$\ \ \ $

\vfill\eject\noindent
%%%%%%%%%%%%%%%%%%%%%%%%%%%%%%%%%%%%%%%%%
          
\narrowtext

%%%%%%%%%%%%%%%%%%%%%%%%%%%%%%%%%%%%%%%%%%%%%%%%%%%%%%%%%%%%%%%%%%%%%%%%%%%%%%
%%%%%%%%%%%%%%%%%%%%%%%%%%%%%%%%%%%%%%%%%%%%%%%%%%%%%%%%%%%%%%%%%%%%%%%%%%%%%%
\section{Hadron tensor and structure functions}
\label{hadron}
\setcounter{equation}{0}

Although the helicity couplings and polarization dependence
are clearly shown in the formalism of the previous section, 
it is not straightforward to find the physics meaning of these 
new structure functions. It is also important to check whether the
number 22 is the right one in an independent way.
A popular method to describe the polarized pp Drell-Yan processes
was developed by Ralston and Soper (RS) \cite{rs} by noting
possible Lorentz index combinations in the hadron
tensor $W^{\mu\nu}$.
The RS type formalism is extended to the reactions of
spin-1/2 and spin-1 hadrons in this section.

We expand the hadron tensor
\begin{equation}
W^{\mu \nu} \! \! = \! \! \!  \int \! \! 
     \frac{d^4 \xi}{(2 \, \pi)^4} \, e^{i Q\cdot \xi} \! \! \!
     < \! \! P_A S_A P_B S_B | J^\mu (0) J^\nu (\xi) | P_A S_A P_B S_B \! \! >
\, ,
\end{equation}
in terms of possible Lorentz index combinations including
the hadron momenta and spins. 
We use the Lorentz vectors $X^\mu$, $Y^\mu$, and $Z^\mu$
in Ref. \cite{rs}:
\begin{align}
X^{\mu}  & =  P_A^{\mu} \, Q^2 \, Z \cdot P_B 
            - P_B^{\mu} \, Q^2 \, Z \cdot P_A
\nonumber \\ \ \ \ \ \ 
&   + Q^{\mu} \, (Q \cdot P_B \, Z \cdot P_A
                          - Q\cdot P_A \, Z \cdot P_B )
\ ,
\nonumber \\
Y^{\mu} & = \epsilon^{\mu \alpha \beta \gamma} 
               \ P_{A \alpha} \, P_{B \beta} \, Q_{\gamma}
\ ,
\nonumber \\
Z^{\mu} & = P_{A}^{\mu} \, Q \cdot P_{B}- P_{B}^{\mu} \, Q \cdot P_A
\ ,
\label{eqn:xyz}
\end{align}
where the convention for the antisymmetric tensor is  $\varepsilon_{0123} =1$.
Then, we define the vector $T^\mu$ as \cite{soper}
\begin{equation}
T^\mu = \varepsilon^{\mu\alpha\beta\gamma} S_{\alpha} \, Z_\beta \, Q_\gamma
\ .
\label{eqn:def-t}
\end{equation}
In addition to the vectors in Eq. (\ref{eqn:xyz}), $Q^\mu$, 
$S_A^\mu$, $S_B^\mu$, $T_A^\mu$, and $T_B^\mu$ are available
for the analysis of $W^{\mu\nu}$.
Instead of $S_A^\mu$ and $S_B^\mu$, it is more convenient to use
the transverse vectors $S_{AT}^{\mu}$ and $S_{BT}^{\mu}$,
which are defined by
\begin{equation}
S_T^{\mu} = \left ( g^{\mu\nu} - \frac{Q^\mu Q^\nu}{Q^2}
                               - \frac{Z^\mu Z^\nu}{Z^2} \right ) S_\nu
\ .
\end{equation}
Even though $S^{\mu}$ is replaced by $S_T^{\mu}$ in Eq. (\ref{eqn:def-t}),
$T^\mu$ remains the same:
$T^\mu = \varepsilon^{\mu\alpha\beta\gamma} S_{\alpha} Z_\beta Q_\gamma
       = \varepsilon^{\mu\alpha\beta\gamma} S_{T \alpha} Z_\beta Q_\gamma$,
and it is a transverse vector $T^\mu = T_T^\mu$.
We should obtain the 108 independent
structure functions in the previous section by the combinations of these
Lorentz vectors and pseudovectors. However, because it is too lengthy to
write them down and all of them are not important in any case, we consider
the limit $Q_T \rightarrow 0$. As we discuss in the end of this section,
the same number of structure functions should be obtained as the one
after integration over $\Phi$ in section \ref{pd-dy}. 
The transverse momentum $Q_T$ is usually small because it is primarily
due to the intrinsic transverse momenta of the partons.
Therefore, it is roughly restricted by the hadron size $r$
as $Q_T \lesssim 1/r$.

In the case $Q_T \rightarrow 0$, the situation becomes simpler because
$|\vec X|$ and $|\vec Y|$ are proportional to $Q_T$ and $X^0=Y^0=0$
in the dilepton rest frame. It means that we do not have to take into
account the vectors $X^\mu$ and $Y^\mu$ in analyzing $W^{\mu\nu}$. 
Furthermore, the hadron tensor has to satisfy the conditions
\begin{align}
& \text{Hermiticity:} 
\nonumber \\ 
&   \ \ 
[W^{\nu \mu}(Q; P_A S_A; P_B S_B)]^* 
             = W^{\mu \nu} (Q; P_A S_A; P_B S_B) \, ,
\nonumber \\  
& \text{parity conservation:} 
\nonumber \\ 
&   \ \  
W^{\mu \nu}(Q; P_A S_A; P_B S_B) =
  W_{\mu \nu}(\overline{Q}; \overline{P}_A -\overline{S}_A; 
                            \overline{P}_B -\overline{S}_B) \, ,
\nonumber \\  
&  \text{time-reveral invariance:}  
\nonumber \\  
&   \ \                
[W^{\mu \nu}(Q; P_A S_A; P_B S_B)]^* = 
         W_{\mu \nu}(\overline{Q}; \overline{P}_A \overline{S}_A; 
                                   \overline{P}_B \overline{S}_B)
\, ,
\label{eqn:hpt}
\end{align}
where $\overline P$ is defined by $\overline P^{\, \mu} = (P^0, -\vec P)$.
Expanding the hadron tensor in terms of all the possible combinations
of the Lorentz vectors and pseudovectors so as to meet these
requirements, we obtain
\begin{align}
W & ^{\mu\nu} = - g^{\mu\nu} A - \frac{Z^\mu Z^\nu}{Z^2} B' 
        + Z^{ \{ \mu} T_A^{\nu \} } C + Z^{ \{ \mu} T_B^{\nu \} } D
\nonumber \\
&  + Z^{ \{ \mu} S_{AT}^{\nu \} } E  + Z^{ \{ \mu} S_{BT}^{\nu \} }  F 
   - S_{BT}^{\mu} S_{BT}^{\nu}    G' - S_{AT}^{ \{ \mu} S_{BT}^{\nu \} } H'
\nonumber \\
&  + T_A^{ \{ \mu} S_{BT}^{\nu \} } I' + S_{BT}^{\{ \mu} T_B^{\nu \} } J 
   + Q^\mu Q^\nu K + Q^{ \{ \mu} Z^{\nu \} } L 
\nonumber \\
&  + Q^{ \{ \mu} S_{AT}^{\nu \} } M   + Q^{ \{ \mu} S_{BT}^{\nu \} } N 
   + Q^{ \{ \mu} T_A^{\nu \} } O   + Q^{ \{ \mu} T_B^{\nu \} } P
\ ,
\label{eqn:w1}
\end{align}
where the notation $Q^{ \{ \mu} Z^{\nu \} }$ is, for example, defined by
\begin{equation}
Q^{ \{ \mu} Z^{\nu \} } \equiv Q^\mu Z^\nu + Q^\nu Z^\mu
\ .
\end{equation}
The coefficients $E$, $F$, $I'$, $J$, $M$, and $N$ are pseudoscalar 
quantities since $S_{AT}^\mu$ and $S_{BT}^\mu$ are pseudovectors. 
We should be careful that the $S_B^2$ type terms are allowed because
the hadron $B$ is a spin-1 particle, whereas only the linear spin
terms are allowed for the hadron $A$. As discussed in section \ref{pd-dy}
repeatedly, the rank-two tensors are possible in $W^{\mu\nu}$.
Because the factors $T_B^\mu T_B^\nu$, $T_A^{ \{ \mu} T_B^{\nu \} }$,
and $S_A^{T \{ \mu} T_B^{\nu \} }$ are not independent from the others,
these are not included in Eq. (\ref{eqn:w1}). This fact could be seen
that the obtained cross sections from these terms have exactly the same
polarization and angular dependence as the other ones.
The hadron tensor $W^{\mu\nu}$ is expressed in terms of the factors
$A$, $B'$, $\cdot\cdot\cdot$, and $P$ as the coefficients.
However, in addition to the conditions in Eq. (\ref{eqn:hpt}),
it has to satisfy the current conservation
\begin{equation}
Q_\mu W^{\mu \nu} = 0 
\ .
\end{equation}
Noting the relations 
$Q\cdot Z = Q\cdot S_{AT} = Q\cdot S_{BT} = Q\cdot T_A = Q\cdot T_B=0$,
we obtain $K=A/Q^2$ and $L=M=N=O=P=0$.
Using these relations and adding the current conserving terms
$g^{\mu\nu} - Q^\mu Q^\nu / Q^2$ and
$g^{\mu\nu} - Q^\mu Q^\nu / Q^2
              - Z^\mu Z^\nu / Z^2$
into the hadron tensor, we have
%%%%%%%%%%%%%%%%%%%%%%%%%%%%%%%%%%%%%%%%%%%%%%%%%%%%%%%%%%%%%%%%%%%%%%%%%%%%%%
\vfill\eject
\widetext
%%%%%%%%%%%%%%%%%%%%%%%%%%%%%%%%%%%%%%%%%%%%%%%%%%%%%%%%%%%%%%%%%%%%%%%%%%%%%%
\begin{align}
W & ^{\mu\nu} = - \left ( g^{\mu\nu} - \frac{Q^\mu Q^\nu}{Q^2} \right ) A 
                - \left [ \frac{Z^\mu Z^\nu}{Z^2} 
                          -\frac{1}{3} \left ( g^{\mu\nu} 
                          - \frac{Q^\mu Q^\nu}{Q^2} \right ) \right ] B
        + Z^{ \{ \mu} T_A^{\nu \} }   C  + Z^{ \{ \mu} T_B^{\nu \} }    D
        + Z^{ \{ \mu} S_{AT}^{\nu \} } E  + Z^{ \{ \mu} S_{BT}^{\nu \} }  F 
\nonumber \\
&  - \left [ S_{BT}^{\mu} S_{BT}^{\nu} 
            - \frac{1}{2} S_{BT} \cdot S_{BT} 
                    \left( g^{\mu\nu} - \frac{Q^\mu Q^\nu}{Q^2}
                          - \frac{Z^\mu Z^\nu}{Z^2} \right ) \right ]   G  
   - \left [ S_{AT}^{ \{ \mu} S_{BT}^{\nu \} } 
            - S_A^T \cdot S_B^T 
                    \left( g^{\mu\nu} - \frac{Q^\mu Q^\nu}{Q^2}
                          - \frac{Z^\mu Z^\nu}{Z^2} \right ) \right ]   H  
\nonumber \\
&  + \left [ T_A^{\{ \mu} S_{BT}^{\nu \} } 
            - T_A \cdot S_{BT} 
                    \left( g^{\mu\nu} - \frac{Q^\mu Q^\nu}{Q^2}
                          - \frac{Z^\mu Z^\nu}{Z^2} \right ) \right ]   I  
   + S_{BT}^{\{ \mu} T_B^{\nu \} } J 
\ .
\label{eqn:w2}
\end{align}
In this way, we expressed the hadron tensor $W^{\mu\nu}$ in terms of
the ten coefficients $A$, $B$, $\cdot\cdot\cdot$, and $J$.
However, these coefficients could still contain
the spin factors in scalar or pseudoscalar forms. 
For example, the coefficient $A$ is obviously a scalar function.
We consider possible scalar combinations among
$1$, $Z$, $S_A$, $S_B$, $T_A$, and $T_B$ up to the order of $S_A$ and $S_B^2$,
then a structure function is assigned to each combination:
\begin{align}
A & = A_1' + \frac{M_A M_B}{s \, Z^2} \, Z\cdot S_A \, Z\cdot S_B \, A_2 
       - S_{AT} \cdot S_{BT} \, A_3
       + \frac{8 \, M_B^2 \, (Z\cdot S_B)^2}{s^2 \, (Q\cdot P_B)^2} A_4'
     + \frac{M_B}{Z^2 \, Q\cdot P_B} \, Z\cdot S_B \, T_A \cdot S_{BT} \, A_5
\nonumber \\
  & = A_1 + \frac{M_A M_B}{s \, Z^2} \, Z\cdot S_A \, Z\cdot S_B \, A_2 
       - S_{AT} \cdot S_{BT} \, A_3
       - \left [ \frac{8 \, M_B^2 \, (Z\cdot S_B)^2}{s^2 \, (Q\cdot P_B)^2}
                 + \frac{4}{3} \, S_B^2  \right ] A_4
     + \frac{M_B}{Z^2 \, Q\cdot P_B} \, Z\cdot S_B \, T_A \cdot S_{BT} \, A_5
\nonumber \\
  & = A_1 + \frac{1}{4} \lambda_A \, \lambda_B \, A_2 
          + |\vec S_{AT}| \, |\vec S_{BT}| \, 
                 cos (\phi_A - \phi_B) \, A_3
          + \frac{2}{3} \, (2 \, |\vec S_{BT}|^2 - \lambda_B^2 ) \, A_4
          + \lambda_B \, |\vec S_{AT}| \, |\vec S_{BT}| \, 
                               sin (\phi_A - \phi_B) \, A_5
\ , 
\label{eqn:a}
\end{align}
where the spin vectors are decomposed into the longitudinal and
transverse components by
\begin{align}
S_A^\mu &= \lambda_A P_A^\mu/M_A + S_{A T}^\mu 
                      - \delta_-^\mu(\lambda_A M_A/P_A^+)
\ ,
\nonumber \\
S_B^\mu &= \lambda_B P_B^\mu/M_B + S_{B T}^\mu  
                      - \delta_+^\mu(\lambda_B M_B/P_B^-)
\ .
\end{align}
The $\lambda_A$ and $\lambda_B$ are the helicities of the hadrons
$A$ and $B$, the momenta $P^+$ and $P^-$ are defined by
$P^\pm =(P^0 \pm P^3)/\sqrt{2}$, and
$\delta_\pm^\mu$ is defined by
$\delta_+^\mu=[0,1,\vec 0_T]$ and $\delta_-^\mu=[1,0,\vec 0_T]$
in the expression of $a^\mu=[a_-,a_+,\vec a_{_T}]$.
The helicity and the transverse vector have the relation,
$\lambda^2+ |\vec S_T|^2=1$.
The $A_1'$ and $A_4'$ terms are combined so that 
the quantity $(Z\cdot S_B)^2$ and the constant $S_B^2=-1$
become a typical quadrupole term $2(2 - 3 \lambda_B^2)/3$.
The $S_B^2$ term is not included in the first line
of Eq. (\ref{eqn:a}) because it is merely a constant.
In obtaining the last line of Eq. (\ref{eqn:a}),
the relations
\begin{align}
\vec Z & = (0,0,| \vec Z|) \ ,
\nonumber \\
\vec S_{AT} & =  |\vec S_{AT}| \, 
                         ( cos \phi_A, sin \phi_A, 0) \ ,
\nonumber \\
\vec S_{BT} & =  |\vec S_{BT}| \, 
                      ( cos \phi_B, sin \phi_B, 0) \ ,
\nonumber \\
\vec T_A        & =  Q \, |\vec Z| \, |\vec S_{AT}| \,
                             ( sin \phi_A, - cos \phi_A, 0) \ ,
\nonumber \\
\vec T_B        & =  Q \, |\vec Z| \, |\vec S_{BT}| \,
                             ( sin \phi_B, - cos \phi_B, 0)
\ ,
\end{align}
are used.
Because the coefficient $B$ is in $W^{\mu\nu}$ of Eq. (\ref{eqn:w2})
without multiplication of any spin factor, it is written
in the same form as Eq. (\ref{eqn:a}):
\begin{align}
B & = B_1 + \frac{M_A M_B}{s \, Z^2} \, Z\cdot S_A \, Z\cdot S_B \, B_2 
       - S_{AT} \cdot S_{BT} \, B_3
       - \left [ \frac{8 \, M_B^2 \, (Z\cdot S_B)^2}{s^2 \, (Q\cdot P_B)^2}
                 + \frac{4}{3} \, S_B^2  \right ] B_4
     + \frac{M_B}{Z^2 \, Q\cdot P_B} \, Z\cdot S_B \, T_A \cdot S_{BT} \, B_5
\nonumber \\
  & = B_1 + \frac{1}{4} \lambda_A \, \lambda_B \, B_2 
          + |\vec S_{AT}| \, |\vec S_{BT}| \, 
                 cos (\phi_A - \phi_B) \, B_3
          + \frac{2}{3} \, ( 2 \, |\vec S_{BT}|^2 - \lambda_B^2 ) \, B_4
          + \lambda_B \, |\vec S_{AT}| \, |\vec S_{BT}| \, 
                                sin (\phi_A - \phi_B) \, B_5
\ .
\end{align}
In the case of the coefficient $C$, the spin factor $S_A$ is already
multiplied through $T_A$ in Eq. (\ref{eqn:w2}).
Therefore, the hadron-$A$ spin factor
cannot be used any longer and it becomes
\begin{equation}
C = - \frac{1}{Q \, Z^2} \, \left [ \, C_1 
    - \left \{ \frac{8 \, M_B^2 \, (Z\cdot S_B)^2}{s^2 \, (Q\cdot P_B)^2}
                 + \frac{4}{3} \, S_B^2  \right \}  C_2 \, \right ]
  = + \frac{1}{Q \, |\vec Z^2|} \, \left [ \, C_1 
             + \frac{2}{3} \, 
               ( 2 \, |\vec S_{BT}|^2 - \lambda_B^2 ) \, C_2 \, \right ]
\ .        
\end{equation}
The factor $1/Q^2 Z^2$ is multiplied so as to cancel the kinematic
factor which appears in calculating 
$(\delta_{ij} -\hat k_i \hat k_j) (Z_i T_{Aj}+Z_j T_{Ai})$
for obtaining the cross section.
%%%%%%%%%%%%%%%%%%%%%%%%%%%%%%%%%%%%%%%%%%%%%%%%%%%%%%%%%%%%%%%%%%%%%%%%%%%%%%
\vfill\eject

$\ \ \ $
\vfill\eject
\widetext
\noindent
%%%%%%%%%%%%%%%%%%%%%%%%%%%%%%%%%%%%%%%%%%%%%%%%%%%%%%%%%%%%%%%%%%%%%%%%%%%%%%
Noting that the coefficient $D$ is multiplied by the spin factor $S_B$,
we have
\begin{align}
D &= - \frac{1}{Q \, Z^2} \, \left [ \, D_1 
  + \frac{M_A \, M_B}{s \, Z^2} \, Z\cdot S_A \, Z\cdot S_B \, D_2
     - S_{AT} \cdot S_{BT} \, D_3  \, \right ]
\nonumber \\
  &= + \frac{1}{Q \, |\vec Z|^2} \, \left [ \, D_1 
          + \frac{1}{4} \, \lambda_A \, \lambda_B \, D_2
          + |\vec S_{AT}| \, |\vec S_{BT}| \, 
                 cos (\phi_A - \phi_B) \, D_3  \, \right ]
\ .        
\end{align}
In the same way, we extract spin dependent factors from the other
coefficients:
\begin{align}
E & =   \frac{Q \, M_B}{Z^2 \, Q\cdot P_B} \, Z\cdot S_B \, E_1
    = - \frac{1}{|\vec Z|} \, \lambda_B \, E_1 \ ,
\nonumber \\
F & = - \frac{Q \, M_A}{Z^2 \, Q\cdot P_A} \, Z\cdot S_A \, F_1
      + \frac{Q \, M_B}{Z^2 \, Q\cdot P_B} \, Z\cdot S_B \, F_2
      - \frac{1}{Z^2 \, Q} \, T_A \cdot S_{BT} \, F_3
\nonumber \\
  & = - \frac{1}{|\vec Z|} \, \left [
           \, \lambda_A \, F_1 + \lambda_B \, F_2
            + |\vec S_{AT}| \, |\vec S_{BT}| \, 
                 sin (\phi_A - \phi_B)  \, F_3  \right ]
\ ,
\nonumber \\
G & = 2 \, G_1 \ , \ \ \ 
   H=H_1 \ , \ \ \
   I=-\frac{M_B}{Z^2 \, Q\cdot P_B} \, Z\cdot S_B \, I_1
    = \frac{\lambda_B}{Q\, |\vec Z|} \, I_1      \ , \ \ \
   J=\frac{M_A}{Z^2 \, Q\cdot P_A} \, Z\cdot S_A \, J_1
    =\frac{\lambda_A}{Q\, |\vec Z|} \, J_1   
\ ,     
\end{align}
where the coefficients $E$, $F$, $I$, and $J$ are pseudoscalar quantities.

Now, we are ready to calculate the cross section by substituting
the coefficients $A$, $B$, $\cdot\cdot\cdot$, and $J$
into Eq. (\ref{eqn:w2}) and then by using Eq. (\ref{eqn:cross0}).
The lepton momentum $\vec k$ is expressed in the polar coordinate as
\begin{equation}
\vec k = |\vec k| \, ( sin \theta \, cos \phi, 
                       sin \theta \, sin \phi, cos \theta )
\ .
\end{equation}
Calculating the cross section, we find redundancy in
the structure functions.
As the coefficient of the $2 \, sin \theta \, cos \theta$ term
in the cross section, we have the factor
\begin{align}
& sin (\phi -\phi_A) \, |\vec S_{AT}| \, 
   [ \, C_1 + 2 \, ( 2 \, |\vec S_{BT}|^2 - \lambda_B^2 ) \, C_2  / 3 \, ] 
\nonumber \\
& \ \ \ \ \ \ \ \ \ \ \ 
  + |\vec S_{AT}| \, |\vec S_{BT}|^2 \, [ \, 
        sin (\phi - \phi_B) \, cos (\phi_A-\phi_B) \, D_3
  +     cos (\phi - \phi_B) \, sin (\phi_A-\phi_B) \, F_3 \, ]
\nonumber \\
= & sin (\phi -\phi_A) \, |\vec S_{AT}| \, [ \, 
             \{ \, C_1 + (D_3 - F_3)/6 \, \} \,
          + 2 \, ( 2 \, |\vec S_{BT}|^2 - \lambda_B^2 ) \, 
             \{ \, C_2 + (D_3 - F_3)/4 \, \}  / 3 \, ]
\nonumber \\
& \ \ \ \ \ \ \ \ \ \ \ 
    + |\vec S_{AT}| \, |\vec S_{BT}|^2 \, sin (\phi+\phi_A-2\phi_B) \, 
        (D_3 + F_3 )/2 
\ .
\end{align}
This equation means that only three of the functions $C_1$, $C_2$, $D_3$,
and $F_3$ are independent because we could redefine the functions as 
$C_1' = C_1 + (D_3 - F_3)/6$,
$C_2' = C_2 + (D_3 - F_3)/4$, and
$D_3' = (D_3  + F_3 )/2$.
This is the same as taking $F_3 = D_3$ without loosing generality, 
therefore we simply use $F_3 = D_3$ in our formalism. 
In this way, the hadron tensor is written
in terms of the 22 structure functions: $A_1$, $A_2$, 
$\cdot\cdot\cdot$, and $J_1$. In order to avoid confusion,
we mention that the functions $F_1$, $F_2$, $F_3$, and $G_1$
are nothing to do with those structure functions in
the lepton-nucleon scattering.

The physics meaning is not clear in
the notations $A_1$, $A_2$, $\cdot\cdot\cdot$, and $J_1$,
so that they are expressed in terms of the notations similar to those
in Refs. \cite{rs,tm}:
\begin{alignat}{5}
A_1=& W_{0,0} \, ,  
               & A_2=& V_{0,0}^{LL} \, , & A_3=& V_{0,0}^{TT} \, , 
               & A_4=& V_{0,0}^{U Q_0} \, , & A_5=& V_{0,0}^{TQ_1} \, ,   
\nonumber \\
B_1=& W_{2,0} \, ,  
               & B_2=& V_{2,0}^{LL} \, , & B_3=& V_{2,0}^{TT} \, ,
               & B_4=& V_{2,0}^{U Q_0} \, , & B_5=& V_{2,0}^{TQ_1} \, ,
\nonumber \\
C_1=& U_{2,1}^{T U} \, , \ \ \ \ \ \ 
                        & C_2=& U_{2,1}^{TQ_0} \, , \ \ \ \ \ \                         & D_1=& U_{2,1}^{U T} \, , \ \ \ \ \ \ 
                        & D_2=& U_{2,1}^{LQ_1} \, , \ \ \ \ \ \ 
                        & D_3=& U_{2,1}^{TQ_2} \, ,
\nonumber \\
E_1=& U_{2,1}^{TL} \, , 
             & F_1=& U_{2,1}^{LT}  \, , 
             & F_2=& U_{2,1}^{U Q_1}  \, , 
             & &   & &
\nonumber \\
G_1=& U_{2,2}^{U Q_2} \, , 
             & H_1=& U_{2,2}^{TT}  \, , 
             & I_1=& U_{2,2}^{TQ_1} \, , 
             & J_1=& U_{2,2}^{LQ_2} \, .  
                         & & 
\end{alignat}
The functions $W$, $V$, and $U$ indicate an unpolarized structure function,
a polarized one without the spin factors $S^\mu$ and $T^\mu$
in Eq. (\ref{eqn:w2}), and a polarized one with the spin factor. 
The subscripts of these structure functions
indicate, for example, that $W_{L,M}$ is obtained by
$\int d\Omega \,  Y_{LM} \, d\sigma/(d^4Q \,d\Omega) \propto W_{L,M}$
in the unpolarized reaction. The superscripts indicate the polarization
states of $A$ and $B$: e.g. $U_{L,M}^{pol_A \, pol_B}$.
The superscripts $U$, $L$, and $T$ indicate unpolarized,
longitudinally polarized, and transversely polarized states.
The quadrupole polarizations $Q_0$, $Q_1$, and $Q_2$ are specific
in the reactions with a spin-1 hadron, and
they are associated with the quadrupole terms in section \ref{pd-dy}:
\begin{alignat}{3}
Q_0   \ \ \ \ \ & \text{for the term} \ \ 
                    & 3 \, cos^2 \beta_B -1 \sim Y_{20} &
\nonumber \\
Q_1   \ \ \ \ \ & \text{for} &  sin \beta_B \, cos \beta_B \sim Y_{21} &
\nonumber \\
Q_2   \ \ \ \ \ & \text{for} &  sin^2 \beta_B \sim Y_{22} &
\ .
\end{alignat}

%%%%%%%%%%%%%%%%%%%%%%%%%%%%%%%%%%%%%%%%%
\vfill\eject

$\ \ \ $

\vfill\eject\noindent
%%%%%%%%%%%%%%%%%%%%%%%%%%%%%%%%%%%%%%%%%

Using these structure-function expressions, we finally obtain
the hadron tensor as
\begin{align}
W^{\mu \nu} =& -\left[ g^{\mu \nu} - \frac{Q^{\mu} Q^{\nu}}{Q^2} \right] 
           \left\{W_{0,0} + \frac{M_A M_B}{S \, Z^2} \,
                        Z \cdot S_A \, Z \cdot S_B \, V_{0,0}^{LL} 
                       - S_{AT} \cdot S_{BT} \, V_{0,0}^{TT} \right.  
\nonumber \\ 
   & \ \ \ \ \ \ \ \ \ \ \ \ 
  - \left(\frac{8 \, M_B^2 \, ( Z \cdot S_B )^2}{S^2 \, ( Q \cdot P_B )^2} 
                             + \frac{4}{3} \, S_B^2 \right) V_{0,0}^{U Q_0} 
                      + \frac{M_B}{Z^2 \, Q \cdot P_B} \,
                              Z \cdot S_B \, T_{A}\cdot S_{BT} \,
                               V_{0,0}^{TQ_1} \biggr\}  
\nonumber \\
                    & -\left[\frac{Z^{\mu} Z^{\nu}}{Z^2}
                      - \frac{1}{3} \left(g^{\mu \nu} 
                         - \frac{Q^{\mu} Q^{\nu}}{Q^2}\right) \right] 
           \left\{W_{2,0} + \frac{M_A M_B}{S \, Z^2} \,
                    Z \cdot S_A \, Z \cdot S_B \, V_{2,0}^{LL}  
                    - S_{AT} \cdot S_{BT} \, V_{2,0}^{TT} \right.  
\nonumber \\
   & \ \ \ \ \ \ \ \ \ \ \ \ 
                     - \left(\frac{8 \, M_B^2 \, ( Z \cdot S_B )^2}
                                {S^2 \, ( Q \cdot P_B )^2} 
                               + \frac{4}{3} \, S_B^2 \right) V_{2,0}^{U Q_0} 
                     + \frac{M_B}{Z^2 \, Q \cdot P_B} \, 
                             Z \cdot S_B \, T_{A} \cdot S_{BT} \, 
                                V_{2,0}^{TQ_1} \biggr\} 
\nonumber \\
                &- Z^{\{\mu} T_{A}^{\nu\}} \frac{1}{\sqrt{Q^2} \, Z^2} 
                     \left\{ U_{2,1}^{T U}
                  - \left( \frac{8 \, M_B^2 \, ( Z \cdot S_B )^2}
                      {S^2 \, ( Q \cdot P_B )^2} 
                        + \frac{4}{3}\, S_B^2 \right) U_{2,1}^{TQ_0} \right\} 
\nonumber \\
                &- Z^{\{\mu} T_{B}^{\nu\}} \frac{1}{\sqrt{Q^2} \, Z^2} 
                      \left\{ U_{2,1}^{U T}
                      + \frac{M_A M_B}{S \, Z^2} \, 
                 Z \cdot S_A  \, Z \cdot S_B \, U_{2,1}^{LQ_1} 
                 + S_{AT} \cdot S_{BT} \, U_{2,1}^{TQ_2} \right\} 
\nonumber \\
                   &+ Z^{\{\mu} S_{AT}^{\nu\}} \,
                      \frac{\sqrt{Q^2} \, M_B}{Z^2 \, Q \cdot P_B} \,
                      Z \cdot S_B \, U_{2,1}^{TL} 
\nonumber \\
                   &+ Z^{\{\mu} S_{BT}^{\nu\}} 
                    \left\{ - \frac{\sqrt{Q^2} \, M_A}{Z^2 \, Q \cdot P_A} \,
                      Z \cdot S_A \, U_{2,1}^{LT}
                      + \frac{\sqrt{Q^2} \, M_B}{Z^2 \, Q \cdot P_B} \, 
                        Z \cdot S_B \, U_{2,1}^{U Q_1} 
                   - \frac{1}{\sqrt{Q^2} \, Z^2} \, T_{A} \cdot S_{BT} \,
                      U_{2, 1}^{TQ_2}\right\} 
\nonumber \\
                &- \left[ 2 \, S_{BT}^{\mu} S_{BT}^{\nu} 
                      -  S_{BT}^2 \left( g^{\mu \nu} 
                               - \frac{Q^{\mu} Q^{\nu}}{Q^2} 
                               -\frac{Z^{\mu} Z^{\nu}}{Z^2} \right) \right]
                     U_{2,2}^{U Q_2}  
\nonumber \\
                &- \left[S_{AT}^{\{\mu} S_{BT}^{\nu\}} 
                    - S_{AT} \cdot S_{BT} \left( g^{\mu \nu} 
                              - \frac{Q^{\mu} Q^{\nu}}{Q^2} 
                              -\frac{Z^{\mu} Z^{\nu}}{Z^2} \right) \right] 
                    U_{2,2}^{TT} 
\nonumber \\
                &- \left[T_{A}^{\{\mu} S_{BT}^{\nu\}} 
                      - T_{A} \cdot S_{BT} \left( g^{\mu \nu} 
                               - \frac{Q^{\mu} Q^{\nu}}{Q^2} 
                               -\frac{Z^{\mu} Z^{\nu}}{Z^2} \right) \right] 
               \frac{M_B}{Z^2 \, Q \cdot P_B} \, Z \cdot S_B \,
                               U_{2,2}^{TQ_1} 
\nonumber \\
                &+ S_{BT}^{\{\mu} T_{B}^{\nu\}} \,
                  \frac{M_A}{Z^2 \, Q \cdot P_A} \, Z \cdot S_A \,
                               U_{2,2}^{LQ_2}
\ .
\end{align}
Substituting this expression into Eq. (\ref{eqn:cross0}),
we obtain the Drell-Yan cross section
\begin{align}
\frac{d\sigma}{d^4Q \, d\Omega} =& 
           \frac{\alpha^2}{2 \, s \, Q^2} \, 
           \left\{ \, 2 \left[ \, W_{0,0}
             + \frac{1}{4} \, \lambda_A \, \lambda_B \, V_{0,0}^{LL}
       + |\vec S_{AT}| \, |\vec S_{BT}| \, \cos (\phi_A - \phi_B) \,
                      V_{0,0}^{TT}
           \right. \right.  
\nonumber \\ 
& \ \ \ \ \ \ \ \ \ 
           + \frac{2}{3} \, \left( 2 \, |\vec S_{BT}|^2 - \lambda_B^2 \right) 
                    \, V_{0,0}^{U Q_0}
           +  |\vec S_{AT}| \, \lambda_B \, |\vec S_{BT}| \, 
                   \sin (\phi_A - \phi_B) \, 
                     V_{0,0}^{TQ_1} \, \biggr] 
\nonumber \\
         &+ \left(\frac{1}{3} - \cos^2 \theta \right) \left[ \, 
                      W_{2,0}
           + \frac{1}{4} \, \lambda_A \, \lambda_B \, V_{2,0}^{LL}
     + |\vec S_{AT}| \, |\vec S_{BT}| \, \cos (\phi_A - \phi_B) \, 
                   V_{2,0}^{TT}
           \right.  
\nonumber \\ 
& \ \ \ \ \ \ \ \ \  
          + \frac{2}{3} \, \left( 2 \, |\vec S_{BT}|^2 - \lambda_B^2 \right)
                    \, V_{2,0}^{U Q_0}
           + |\vec S_{AT}| \, \lambda_B \, |\vec S_{BT}| \, 
                   \sin (\phi_A - \phi_B) \, 
                         V_{2,0}^{TQ_1} \, \biggr] 
\nonumber \\ 
          &+ 2 \sin \theta \cos \theta
    \left[ \,  \sin (\phi - \phi_A) \, |\vec S_{AT}| \, 
                  \left(   \, U_{2,1}^{T U} 
         + \frac{2}{3} \, \left( 2 \, |\vec S_{BT}|^2 - \lambda_B^2 \right)
                       \, U_{2,1}^{TQ_0}\right)
            \right.
\nonumber \\ 
& \ \ \ \ \ \ \ \ \ 
           +  \sin (\phi - \phi_B) \, |\vec S_{BT}| 
                  \left(  U_{2,1}^{U T} 
      + \frac{1}{4} \, \lambda_A \, \lambda_B  \, U_{2,1}^{LQ_1}\right) 
      + \sin (\phi + \phi_A - 2 \phi_B) \,
              |\vec S_{AT}| \, |\vec S_{BT}|^2 \, U_{2,1}^{TQ_2} 
\nonumber \\ 
& \ \ \ \ \ \ \ \ \ 
        + \cos (\phi - \phi_A) \, |\vec S_{AT}| \, \lambda_B \,
            U_{2,1}^{TL}  
        + \cos (\phi - \phi_B) \, |\vec S_{BT}| \, 
        \left(\lambda_A \, U_{2,1}^{LT} 
            + \lambda_B \, U_{2,1}^{U Q_1}\right)
                                     \,   \biggr] 
\nonumber \\ 
          &+ \sin^2 \theta
                 \left[ \, 
                  \cos (2 \phi - 2 \phi_B) \,
                 |\vec S_{BT}|^2 \, U_{2,2}^{U Q_2} 
           + \cos (2 \phi - \phi_A - \phi_B) \,
               |\vec S_{AT}| \, |\vec S_{BT}| \, U_{2,2}^{TT} \right.
\nonumber \\ 
& \ \ \ \ \ \ \ \ \ 
            + \, \sin (2 \phi - \phi_A - \phi_B) \,
            |\vec S_{AT}| \,\lambda_B \, |\vec S_{BT}| 
                      \, U_{2,2}^{T Q_1} 
            + \, \sin (2 \phi - 2 \phi_B) \,
                 \lambda_A \, |\vec S_{BT}|^2 \, 
                         U_{2,2}^{LQ_2} \, 
             \biggr] \,  \biggr\} 
\ .   
\label{eqn:cross-w}                            
\end{align}

%%%%%%%%%%%%%%%%%%%%%%%%%%%%%%%%%%%%%%%%%
\vfill\eject

$\ \ \ $

\vfill\eject
\narrowtext
\noindent
%%%%%%%%%%%%%%%%%%%%%%%%%%%%%%%%%%%%%%%%%     
Using $\lambda_A=|\vec S_A| cos \theta_A$, 
      $\lambda_B=|\vec S_B| cos \theta_B$,
      $|\vec S_{AT}| = |\vec S_A| sin \theta_A$,
  and $|\vec S_{BT}| = |\vec S_B| sin \theta_B$
in Eq. (\ref{eqn:cross-w}), and noting the relations:
$\theta_A=\beta_A$, $\phi_A=\alpha_A$,
$\theta_B=\pi-\beta_B$, and $\phi_B=-\alpha_B$,
we obtain essentially the same equation as Eq. (\ref{eqn:cross4}).
In this way, we find that 22 structure functions exist
in the Drell-Yan processes with spin-1/2 and spin-1 hadrons
in the limit $Q_T\rightarrow 0$.
We can repeat the analysis of this section by taking the integral
over $\vec Q_T$ first. However, $X^\mu$ and $Y^\mu$ are averaged away
by the $\vec Q_T$ integration. It means that exactly the same formalism
can be applied without taking into account the vectors $X^\mu$ and $Y^\mu$ 
except for the replacements \cite{rs}, 
$W_{00}\rightarrow \overline W _{00}=\int d \vec Q_T W _{00}$,
$W_{20}\rightarrow \overline W _{20}=\int d \vec Q_T W _{20}$, 
and others. This is the reason why 
the same number of the structure
functions are obtained after taking the limit $Q_T\rightarrow 0$
as discussed in this section and after the integration over $\vec Q_T$
(or $\Phi$) as discussed in section \ref{pd-dy}.

Because there exist 11 structure functions after the $\vec Q_T$
integration in the Drell-Yan processes of spin-1/2 hadrons,
there are 11 new structure functions in the reactions
of spin-1/2 and spin-1 hadrons.
In the pp Drell-Yan processes, the structure functions
$U_{2,1}^{TL}$ and $U_{2,1}^{UT}$ are identified with 
$U_{2,1}^{LT}$ and $U_{2,1}^{TU}$. 
Therefore, there exist essentially 9 structure functions.
In this sense, there are 13 additional structure functions in our
Drell-Yan processes.  As it is shown in Eq. (\ref{eqn:cross-w}), 
the 11 new structure functions are related to the quadrupole
polarizations $Q_0$, $Q_1$, and $Q_2$. In order to measure
these structure functions, various longitudinal, transverse,
and intermediate polarization reactions should be combined.
These are key structure functions to
clarify the tensor structure of the spin-1 hadrons.

%%%%%%%%%%%%%%%%%%%%%%%%%%%%%%%%%%%%%%%%%%%%%%%%%%%%%%%%%%%%%%%%%%%%%%%%%%%%%%
%%%%%%%%%%%%%%%%%%%%%%%%%%%%%%%%%%%%%%%%%%%%%%%%%%%%%%%%%%%%%%%%%%%%%%%%%%%%%%
\section{Spin asymmetries}
\label{asym}
\setcounter{equation}{0}

In the previous sections, we have derived the expressions
for the Drell-Yan cross sections in Eqs. (\ref{eqn:cross4})
and (\ref{eqn:cross-w}). However, it is not straightforward
to see how each structure function can be measured.
In this section, we discuss the relations between the
structure functions and possible spin asymmetries.
Because the Ralston-Soper type notations seem to be
more popular, we use the cross-section expression 
in Eq. (\ref{eqn:cross-w}) for discussing the spin asymmetries.

In the proton-proton Drell-Yan cross sections,
there are merely the following spin combinations \cite{rs,dg,single}:
\begin{equation}
< \! \sigma \! >, \ \ 
A_{LL}, \ \ 
A_{TT}, \ \ 
A_{LT}, \ \ 
A_{T} \ \ \ \text{(in pp)}
\ ,
\end{equation}
where 
$< \! \sigma \! >$ is the unpolarized cross section,
$A_{LL}$ ($A_{TT}$) is the longitudinal (transverse) double spin asymmetry,
$A_{LT}$ is the longitudinal-transverse spin asymmetry,
and $A_T$ is the transverse single spin asymmetry.
Here, the parity-violating asymmetries are not
taken into account \cite{parity-pp}. 
As it is obvious from the cross section in Eq. (\ref{eqn:cross-w}),
there are many other spin combinations in the reactions
with spin-1/2 and spin-1 hadrons (proton and deuteron).

First, the expression of our unpolarized cross section is the same
as the pp one:
\begin{equation}
\left < \frac{d\sigma}{d^4Q \, d\Omega} \right> =
         \frac{\alpha^2}{2 \, s \, Q^2} \, 
         \left [ \, 2 \, W_{0,0}
                 + \left ( \frac{1}{3} - cos^2\theta \right ) \, 
                         W_{2,0} \, \right ] 
\ .
\end{equation}
The longitudinal double asymmetry may be defined in the similar
way as the pp case: 
$A_{LL}'=[ \sigma(\uparrow , -1) -
           \sigma(\uparrow , $ $  +1) ] /
        [ \sigma(\uparrow , -1) + \sigma(\uparrow , +1) ]$.
However, this is not an appropriate way to investigate the
longitudinally polarized parton distributions because the
tensor distribution $b_1$ contributes to the asymmetry
$A_{LL}$. This asymmetry definition is used in analyzing
the polarized lepton-deuteron scattering data for extracting
$g_1^n$. Strictly speaking, this is not an appropriate way of
handling the data. In order to avoid this kind of confusion, we use
$2 < \! \sigma \! >$ in the denominator for defining the spin
asymmetry. From the relation
\begin{multline}
2 < \! \sigma \! > \, =  \sigma(\uparrow , +1) + \sigma(\uparrow , -1) 
\\
               + \frac{1}{3} \, 
                [ \, 2 \, \sigma(\uparrow , 0)
                  - \sigma(\uparrow , +1) - \sigma(\uparrow , -1) \, ]
\ ,
\end{multline}
the term $2 < \! \sigma \! >$ agrees with the usual denominator
$\sigma(\uparrow -1) + \sigma(\uparrow +1)$
if the quadrupole asymmetry can be ignored:
$ \left | \, 2 \, \sigma(\uparrow , 0)
         - \sigma(\uparrow , -1) - \sigma(\uparrow , +1) \, \right | 
  \ll \, 6 < \! \sigma \! >$.
In this way, we define a {\it modified} spin asymmetry
by using the denominator $2 < \! \sigma \! >$. Then, the longitudinal
double (LL) spin asymmetry becomes
\begin{align}
A_{LL} & = \frac{ \sigma(\uparrow_L , -1_L) 
                       - \sigma(\uparrow_L , +1_L) }
                    { 2 < \! \sigma \! > }
\nonumber \\
              & = - \frac{ 2 \, V_{0,0}^{LL} 
                          + (\frac{1}{3}-cos^2 \theta ) \, 
                            V_{2,0}^{LL} }
                 { 4 \, \left[ \, 2 \, W_{0,0}
                          + (\frac{1}{3}-cos^2 \theta ) \, 
                            W_{2,0} \, \right] }
\ ,
\label{eqn:a-ll}
\end{align}
where the subscripts of $\uparrow_L$, $+1_L$, and $-1_L$
indicate the longitudinal polarization.
Experimentally, it may be easier to obtain the usual asymmetry $A_{LL}'$
rather than the above one. Because the quadrupole effects are considered
to be small, the result would not be changed significantly even if
our $A_{LL}$ is simply replaced by $A_{LL}'$. 
However, the tensor structure is the major point 
in studying the polarized spin-1 hadron, 
so that its contributions to $A_{LL}'$ should be carefully taken
into account. If the cross section is integrated over $\vec Q_T$,
the denominator of Eq. (\ref{eqn:a-ll}) is expressed in terms of
the unpolarized quark and antiquark distributions $f_1$ and
$\overline f_1$ as
$ 2 \,  {\overline W}_{0,0} 
  + (\frac{1}{3}-cos^2 \theta ) \, {\overline W}_{2,0}
  =  (1+cos^2 \theta) \, {\overline W}_T = (1+cos^2 \theta) \, 
       (1/3) \sum_a e_a^2 f_1 \overline f_1$
in a parton model \cite{tm}.
Our parton-model analysis of the reactions with spin-1/2 and
spin-1 hadrons will be reported in Ref. \cite{our2}. 

In the LL asymmetry, the structure-function expression
of Eq. (\ref{eqn:a-ll}) is not altered even if the spin asymmetry
is taken in the hadron-$A$: 
$[\sigma(\downarrow_L , +1_L)
 - \sigma(\uparrow_L ,  +1_L) ] $ $
       /(2 < \! \sigma \! > )$.
However, the situation is different in the transverse double spin
asymmetry. If the asymmetry is taken in the hadron-$B$ with a fixed
transverse polarization of the hadron-$A$,
the transverse double (TT) spin asymmetry becomes
\begin{align}
A_{(T)T} & = \frac{ \sigma(\phi_A=0 , \phi_B=0) 
                         - \sigma(\phi_A=0 , \phi_B=\pi) }
                    { 2 < \! \sigma \! > }
\nonumber \\
               = & \bigg [ \, 2 \, V_{0,0}^{TT} 
                          + (\frac{1}{3}-cos^2 \theta ) \, V_{2,0}^{TT}
                          + sin^2 \theta \, cos 2\phi \,   U_{2,2}^{TT}
\nonumber \\
                    & \! \! \! \! \! \! \! \! \! \! \! \! \! \! \! \! \! 
                          + 2 \, sin \theta \, cos \theta \, sin \phi \,
                              U_{2,1}^{UT} \, 
                  \bigg ] \, \bigg / \left [ \, 2 \, W_{0,0}
                          + (\frac{1}{3}-cos^2 \theta ) \, W_{2,0} \, \right ]
\, .
\label{eqn:a-ttb}
\end{align}
Hereafter, if $\phi_A$ or $\phi_B$ is indicated in the expression
of $\sigma(pol_A,pol_B)$, it means that the hadron $A$ or $B$ is
transversely polarized with the azimuthal angle $\phi_A$ or $\phi_B$.
It should be noted that the transverse asymmetry is not equal to
Eq. (\ref{eqn:a-ttb}) if the spin asymmetry is taken in the hadron-$A$:
\begin{align}
A_{T(T)}  = & \frac{ \sigma(\phi_A=0 , \phi_B=0) 
                         - \sigma(\phi_A=\pi , \phi_B=0) }
                    { 2 < \! \sigma \! > }
\nonumber \\
               = & \bigg [ \, 2 \, V_{0,0}^{TT} 
                          + (\frac{1}{3}-cos^2 \theta ) \, V_{2,0}^{TT}
                          + sin^2 \theta \, cos 2\phi \,   U_{2,2}^{TT}
\nonumber \\
                    &    + 2 \, sin \theta \, cos \theta \, sin \phi \,
                            \left ( U_{2,1}^{TU}
                                    + \frac{4}{3} \, U_{2,1}^{TQ_0}
                                                +  U_{2,1}^{TQ_2} \right ) \, 
                  \bigg ]
\nonumber \\
                  &  \bigg / \left [ \, 2 \, W_{0,0}
                          + (\frac{1}{3}-cos^2 \theta ) \, W_{2,0} \, \right ]
\ .
\end{align}
It is cumbersome to handle the above transverse-transverse asymmetries
$A_{(T)T}$ and $A_{T(T)}$ because other structure functions mix with
the TT-type ones. In order to separate the TT type, the following
parallel transverse-transverse asymmetry should be studied:
\begin{align}
A_{TT}^{\, \parallel} & = \frac{1}{2 < \! \sigma \! >} \,        
         \bigg [ \, 
             \frac{ \sigma(\phi_A=0 , \phi_B=0) 
                         + \sigma(\phi_A=\pi , \phi_B=\pi) }{2}
\nonumber \\
           & \ \ \ \ \ \ \ \ \ 
            - \frac{ \sigma(\phi_A=\pi , \phi_B=0) 
                     + \sigma(\phi_A=0 , \phi_B=\pi) }{2} \, \bigg ]   
\nonumber \\
              & = \frac{ 2 \, V_{0,0}^{TT} 
                          + (\frac{1}{3}-cos^2 \theta ) \, V_{2,0}^{TT}
                          + sin^2 \theta \, cos 2\phi \,   U_{2,2}^{TT} }
                 { 2 \, W_{0,0}
                          + (\frac{1}{3}-cos^2 \theta ) \, W_{2,0} }
\ .
\end{align}
Furthermore, the $U_{2,2}^{TT}$ part can be separated by
the perpendicular transverse-transverse asymmetry
\begin{align}
A_{TT}^{\, \perp} & =      
         \bigg [ \, 
             \frac{ \sigma(\phi_A=0 , \phi_B=\pi/2) 
                         + \sigma(\phi_A=\pi , \phi_B=3 \pi/2) }{2}
\nonumber \\
           & \ \ 
            - \frac{ \sigma(\phi_A=\pi , \phi_B=\pi/2) 
                     + \sigma(\phi_A=0 , \phi_B=3 \pi/2) }{2} \, \bigg ]   
\nonumber \\
              & \bigg / \, (2 < \! \sigma \! >) \,   
= \frac{ sin^2 \theta \, sin 2\phi \,   U_{2,2}^{TT} }
                 { 2 \, W_{0,0}
                          + (\frac{1}{3}-cos^2 \theta ) \, W_{2,0} }
\ .
\label{eqn:attperp}
\end{align}

The situation is also complicated, for example, in the 
longitudinal-transverse (LT) asymmetry in the sense that
different types of the structure functions could contribute
if the asymmetry is taken either in the hadron-$A$ or in $B$.
The optimum  LT and TL asymmetries are given by
\begin{align}
A_{LT} = & \frac{1}{2 < \! \sigma \! >} \,        
         \bigg [ \, 
             \frac{ \sigma(\uparrow_L , \phi_B=0) 
                         + \sigma(\downarrow_L , \phi_B=\pi) }{2}
\nonumber \\
           & \ \ \ \ \ \ \ \ \ \ \ 
            - \frac{ \sigma(\uparrow_L ,  \phi_B=\pi) 
                   + \sigma(\downarrow_L ,\phi_B=0) }{2} \, \bigg ]   
\nonumber \\
              & =  \frac{ 2 \, sin\theta \, cos\theta \, cos\phi \,
                           U_{2,1}^{LT} }
                 { 2 \, W_{0,0}
                          + (\frac{1}{3}-cos^2 \theta ) \, W_{2,0} }
\ ,
\end{align}
\begin{align}
A_{TL} = & \frac{1}{2 < \! \sigma \! >} \,        
         \bigg [ \, 
             \frac{ \sigma(\phi_A=0 , +1_L) 
                         + \sigma(\phi_A=\pi , -1_L) }{2}
\nonumber \\
           & \ \ \ \ \ \ \ \ \ \ \ 
            - \frac{ \sigma(\phi_A=0 ,  -1_L) 
                   + \sigma(\phi_A=\pi ,+1_L) }{2} \, \bigg ]   
\nonumber \\
              & =  \frac{ 2 \, sin\theta \, cos\theta \, cos\phi \,
                           U_{2,1}^{TL} }
                 { 2 \, W_{0,0}
                          + (\frac{1}{3}-cos^2 \theta ) \, W_{2,0} }
\ ,
\label{eqn:atl}
\end{align}
by combining both spin asymmetries in the hadrons $A$ and $B$. 
In the case of the asymmetry $A_{TL}$, we could have
defined it in a simpler form as
$A_{TL} =A_{(T)L} = [\sigma(\phi_A=0 , +1_L) 
    - \sigma(\phi_A=0 , -1_L) ]/ (2 < \! \sigma \! >)$, which becomes
the same equation as Eq. (\ref{eqn:atl}) in terms of
the structure functions $U_{2,1}^{TL}$, $W_{0,0}$, and $W_{2,0}$.
However, we should be careful about the definitions of the LT and TL
asymmetries. If the asymmetries are calculated by
$A_{(L)T} = [ \sigma(\uparrow_L , \phi_B=0) 
                        - \sigma(\uparrow_L , \phi_B=\pi) ]
/(2 < \! \sigma \! >)$,
$A_{L(T)} = [ \sigma(\uparrow_L ,   \phi_B=0) 
                         - \sigma(\downarrow_L , \phi_B=0) ]
/(2 < \! \sigma \! >)$,
and 
$A_{T(L)} = [ \sigma(\phi_A=0   , +1_L) 
                         - \sigma(\phi_A=\pi , +1_L) ]
/(2 < \! \sigma \! >)$,
there are other contributions in addition to $U_{2,1}^{LT}$ or
$U_{2,1}^{TL}$. Therefore, the asymmetries $A_{(L)T}$, $A_{L(T)}$,
and $A_{T(L)}$ are not appropriate quantities for studying exclusively
the LT and TL structure functions.
In the following discussions of this section on various asymmetries,
the details are no longer discussed about a number of different
possibilities. We simply show the optimum asymmetries.

In the single spin asymmetries, we express the unpolarized state explicitly,
for example, as $A_{UT}$ which is the transverse single spin asymmetry
for the hadron $B$. It is because $A_{UT}$ and $A_{TU}$ are in general
different in our case although it does not matter in the pp.
Then, the single spin asymmetries become
\begin{align}
A_{UT} & = \frac{ \sigma(\bullet , \phi_B=0) 
                       - \sigma(\bullet , \phi_B=\pi) }
                    { 2 < \! \sigma \! > }
\nonumber \\
              & =  \frac{ 2 \, sin\theta \, cos\theta \, sin\phi \,
                           U_{2,1}^{UT} }
                 { 2 \, W_{0,0}
                          + (\frac{1}{3}-cos^2 \theta ) \, W_{2,0} }
\ ,
\end{align}
\begin{align}
A_{TU} & = \frac{ \sigma(\phi_A=0  , \bullet) 
                       - \sigma(\phi_A=\pi , \bullet) }
                    { 2 < \! \sigma \! > }
\nonumber \\
              & =  \frac{ 2 \, sin\theta \, cos\theta \, sin\phi \,
                           U_{2,1}^{TU} }
                 { 2 \, W_{0,0}
                          + (\frac{1}{3}-cos^2 \theta ) \,  W_{2,0} }
\ ,
\end{align}
where $\bullet$ indicates the unpolarized case.

\begin{figure}[h]
   \begin{center}
       \epsfig{file=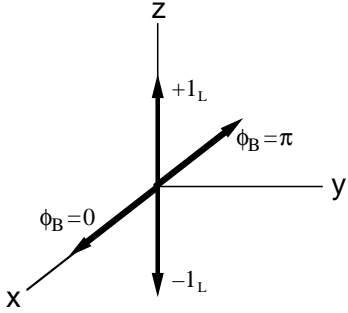,width=5.0cm}
   \end{center}
   \vspace{-0.2cm}
   \caption{Quadrupole asymmetry $Q_0$. The arrows indicate polarization
            directions of the hadron $B$.}
   \label{fig:q0}
\end{figure}

In defining the quadrupole spin asymmetry $Q_0$, we consider
the parton-model definition of the $b_1$ structure function
\cite{mit-b1}: $b_1 = [q_0 -(q_{+1}+q_{-1})/2]/2$ where
the subscripts denote $z$ components of the hadron spin.
The quadrupole asymmetry $Q_0$ is, therefore, defined by
\begin{align}
A_{UQ_0} & = \frac{1}{2 < \! \sigma \! >} \,        
         \bigg [ \, \sigma(\bullet , 0_L)
            - \frac{ \sigma(\bullet , +1_L) 
                    +\sigma(\bullet , -1_L) }{2} \, \bigg ]   
\nonumber \\
              & =  \frac{ 2 \, V_{0,0}^{UQ_0} 
                          + (\frac{1}{3}-cos^2 \theta ) \, 
                            V_{2,0}^{UQ_0} }
                 { 2 \, W_{0,0}
                       + (\frac{1}{3}-cos^2 \theta ) \,  W_{2,0}  }
\ ,
\label{eqn:a-uq0}
\end{align}
where the average is taken over the angle $\phi_B$ in
the polarization state $0_L$. If we wish to specify the azimuthal
angle in defining the asymmetry, it could be written as
$A_{UQ_0}=[ \, \{ \sigma(\bullet , \phi_B=0) 
                         + \sigma(\bullet , \phi_B=\pi) \} /2
            -  \{ \sigma(\bullet , +1_L) 
                    +\sigma(\bullet , -1_L) \} /2 \, ] 
             / (2 < \! \sigma \! >)$.
The situation is illustrated in Fig. \ref{fig:q0}.
The $Q_0$ polarization is related to the difference
between the longitudinal and transverse cross sections.
\vspace{-0.0cm}
%%%%%%%%%%%%%%%%%%%%%%%%%%%%%%%% figure %%%%%%%%%%%%%%%%%%%%%%%%%%%%%%%%%%%%%%
\noindent
%%%%%%%%%%%%%%%%%%%%%%%%%%%%%%%% figure %%%%%%%%%%%%%%%%%%%%%%%%%%%%%%%%%%%%%%
\noindent
This is a new asymmetry which does not exist in the pp reactions.
We will show in Ref. \cite{our2} that this asymmetry is related
to the $b_1$ distribution. This quadrupole asymmetry is one of
the interesting quantities to be investigated in our reactions.
The $b_1$ structure function will be studied in the polarized lepton-deuteron
scattering by the HERMES collaboration \cite{hermes}; 
however, the experimental accuracy may not be good enough
to find the small quantity. Therefore, the idea is to use an accelerator
with enough beam intensity. A realistic possibility is to measure it
in the ELFE (Electron Laboratory for Europe) project \cite{elfe}
or in a similar one \cite{rcnp}. However, these projects
are not approved yet. It would take several years
to start investigating $b_1$ experimentally by considering the present
situation. The polarized pd reactions are the timely and pioneering works
if they are studied in the next generation RHIC-Spin project. 

In the same way, transverse-quadrupole $Q_0$ ($TQ_0$) asymmetry is
given by
\begin{align}
A_{TQ_0} = & \frac{1}{2 < \! \sigma \! >} \,        
         \bigg [ \, 
             \{ \, \sigma(\phi_A=0 , 0_L) 
                         - \sigma(\phi_A=\pi , 0_L) \, \}/2
\nonumber \\
           & \ \ \ \ \ \ \ \  
            - \{ \, \sigma(\phi_A=0 , +1_L) + \sigma(\phi_A=0 , -1_L)
\nonumber \\
           & \ \ \ \ \ \ \ \ 
            - \sigma(\phi_A=\pi , +1_L) - \sigma(\phi_A=\pi , -1_L) \, \}/4 
                   \, \bigg ]   
\nonumber \\
              = & \frac{ 2 \, sin\theta \, cos\theta \, sin\phi \,
                           U_{2,1}^{TQ_0} }
                 {  2 \, W_{0,0}
                          + (\frac{1}{3}-cos^2 \theta ) \, W_{2,0} }
\ .
\end{align}
Here, the $\sigma(\phi_A, 0_L)$ term cannot be replaced by
$ [ \sigma(\phi_A, \phi_B=0) + \sigma(\phi_A, \phi_B=\pi) ] /2$
as it was discussed just after Eq. (\ref{eqn:a-uq0}).
If we would like to specify $\phi_B$, it should be replaced by
$ [ \sigma(\phi_A, \phi_B=0)   + \sigma(\phi_A, \phi_B=  \pi/2) 
   +\sigma(\phi_A, \phi_B=\pi) + \sigma(\phi_A, \phi_B=3 \pi/2) ] /4$
in order to cancel out the $Q_2$-type structure functions.

\begin{figure}[h]
   \begin{center}
       \epsfig{file=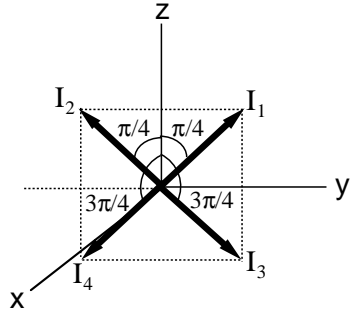,width=5.0cm}
   \end{center}
   \vspace{-0.2cm}
   \caption{Quadrupole asymmetry $Q_1$. The polarization vectors
               are in the $yz$ plane.
}
   \label{fig:q1}
\end{figure}

We find other peculiar asymmetries and structure functions.
There are structure functions which do not exist in the transverse
and longitudinal polarization reactions; however, they can exist
in the intermediate polarization reactions in Fig. \ref{fig:q1}.
\vspace{-0.0cm}
%%%%%%%%%%%%%%%%%%%%%%%%%%%%%%%% figure %%%%%%%%%%%%%%%%%%%%%%%%%%%%%%%%%%%%%%
\noindent
%%%%%%%%%%%%%%%%%%%%%%%%%%%%%%%% figure %%%%%%%%%%%%%%%%%%%%%%%%%%%%%%%%%%%%%%
\noindent
The single spin asymmetry for the intermediate polarization is:
\begin{align}
A_{UQ_1} & = \frac{ \sigma(\bullet , I_1) - \sigma(\bullet , I_3) }
                    { 2 < \! \sigma \! > }
\nonumber \\
              & =  \frac{ sin\theta \, cos\theta \, 
                             sin \phi \, U_{2,1}^{UQ_1} }
                 { 2 \, W_{0,0} 
                          + (\frac{1}{3}-cos^2 \theta ) \, W_{2,0} }
\ ,
\label{eqn:a-uq1}
\end{align}
where the intermediate polarizations are denoted as $I_1$ and $I_3$
and they are shown in Fig. \ref{fig:q1}.
The longitudinal-quadrupole $Q_1$ (or intermediate) asymmetry is
\begin{align}
A_{LQ_1} & = \frac{1}{8 < \! \sigma \! >} \,    
         \bigg [ \, \sigma(\downarrow_L , I_1) 
                   -\sigma(\downarrow_L , I_2)
                   -\sigma(\downarrow_L , I_3)
\nonumber \\
               &   \! \! \! \! \! \! \! \! \! \! 
                   \! \! \! \! \! \! 
                   +\sigma(\downarrow_L , I_4)
                   -\sigma(\uparrow_L , I_1)    
                   +\sigma(\uparrow_L , I_2)    
                   +\sigma(\uparrow_L , I_3)    
                   -\sigma(\uparrow_L , I_4)  \, \bigg ]  \,
\nonumber \\
              & =  \frac{  sin\theta \, cos\theta \, 
                           cos\phi \, U_{2,1}^{LQ_1} }
                 { 4 \, \left[ \, 2 \, W_{0,0}
                          + (\frac{1}{3}-cos^2 \theta ) \, 
                            W_{2,0} \, \right] }
\ ,
\label{eqn:a-lq1}
\end{align}
where the additional polarizations $I_2$ and $I_4$ are also shown
in Fig. \ref{fig:q1}. The asymmetry definition is more complicated
than the one in Eq. (\ref{eqn:a-uq1}) in order to cancel out
other contributions. In the same way, these intermediate polarizations should
be combined with the transverse polarization states of the hadron-$A$
for obtaining the transverse-quadrupole $Q_1$ (or intermediate) asymmetry:
\begin{align}
A_{TQ_1} = & \frac{1}{8 < \! \sigma \! >} \,        
         \bigg [ \, - \sigma(\phi_A=0 ,   I_1) 
                    + \sigma(\phi_A=0 ,   I_2)
\nonumber \\
           & 
                    + \sigma(\phi_A=\pi , I_1)
                    - \sigma(\phi_A=\pi , I_2)
                    + \sigma(\phi_A=0 ,   I_3)
\nonumber \\
           & 
                    - \sigma(\phi_A=0 ,   I_4)
                    - \sigma(\phi_A=\pi , I_3)
                    + \sigma(\phi_A=\pi , I_4)  \, \bigg ]   
\nonumber \\
               =  & \frac{ 2 \, V_{0,0}^{TQ_1} 
                          + (\frac{1}{3}-cos^2 \theta ) \, V_{2,0}^{TQ_1} 
                          +  sin^2 \theta \, cos 2\phi \, U_{2,2}^{TQ_1} }
                 { 2 \, \left[ \, 2 \, W_{0,0}
                          + (\frac{1}{3}-cos^2 \theta ) \, 
                            W_{2,0} \, \right] }
\ .
\label{eqn:atq1}
\end{align}

\begin{figure}[h]
   \begin{center}
       \epsfig{file=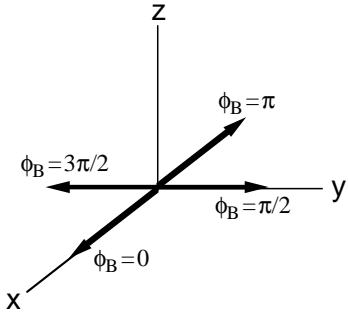,width=5.0cm}
   \end{center}
   \vspace{-0.2cm}
   \caption{Quadrupole asymmetry $Q_2$.}
   \label{fig:q2}
\end{figure}

There are other interesting asymmetries which are related to
the quadrupole polarization in the transverse plane.
There are structure functions with the $sin(2\phi_B)$ or
$cos(2\phi_B)$ factor in Eq. (\ref{eqn:cross-w}).
These terms vanish if the cross-section difference is taken between
those of the opposite transverse polarizations,
$\phi_B=0$ and $\phi_B=\pi$. They are associated
with the difference between $\phi_B=0$ and $\phi_B=\pi/2$
cross sections as illustrated in Fig. \ref{fig:q2}.
\vspace{-0.0cm}
%%%%%%%%%%%%%%%%%%%%%%%%%%%%%%%% figure %%%%%%%%%%%%%%%%%%%%%%%%%%%%%%%%%%%%%%
\noindent
%%%%%%%%%%%%%%%%%%%%%%%%%%%%%%%% figure %%%%%%%%%%%%%%%%%%%%%%%%%%%%%%%%%%%%%%
\noindent
The single quadrupole asymmetry $Q_2$ is defined and it is expressed
in terms of our structure functions as
\begin{align}
A_{UQ_2} & = \frac{1}{2 < \! \sigma \! >} \,        
         \bigg [ \, 
             \frac{ \sigma(\bullet , \phi_B=0) 
                         + \sigma(\bullet , \phi_B=\pi) }{2}
\nonumber \\
           & \ \ \ \ \ \ \ \ \ \ \ 
            - \frac{ \sigma(\bullet , \phi_B=\pi/2) 
                   + \sigma(\bullet , \phi_B=3\pi/2) }{2} \, \bigg ]   
\nonumber \\
              & =  \frac{  sin^2 \theta \, cos 2 \phi \,
                           U_{2,2}^{U Q_2} }
                 { 2 \, W_{0,0} 
                          +  \, (\frac{1}{3}-cos^2 \theta ) \, W_{2,0} }
\label{eqn:a-uq2}
\ .
\end{align}
In the same way, the other $Q_2$ asymmetries are given as
\begin{align}
A_{LQ_2} & = \frac{1}{8 < \! \sigma \! >} \,    
         \bigg [ \, \sigma(\uparrow_L , \phi_B=0) 
                   +\sigma(\uparrow_L , \phi_B=\pi)
\nonumber \\
               &   \! \! \! \! \! \! \! \! \! \! \! \! \! \! \! 
                   -\sigma(\uparrow_L , \phi_B=\pi/2)
                   -\sigma(\uparrow_L , \phi_B=3\pi/2)
                   -\sigma(\downarrow_L , \phi_B=0)    
\nonumber \\  
               &   \! \! \! \! \! \! \! \! \! \! \! \! \! \! \! 
                   -\sigma(\downarrow_L , \phi_B=\pi) 
                   +\sigma(\downarrow_L , \phi_B=\pi/2)    
                   +\sigma(\downarrow_L , \phi_B=3\pi/2)  \, \bigg ] 
\nonumber \\
              & =  \frac{  sin^2 \theta \, sin 2 \phi \,
                           U_{2,2}^{L Q_2} }
                 { 2 \, W_{0,0}
                          +  \, (\frac{1}{3}-cos^2 \theta ) \, W_{2,0} }
\ ,
\end{align}
\begin{align}
A_{TQ_2} & = \frac{1}{8 < \! \sigma \! >} \,      
              \bigg [ \, \sigma(\phi_A=0 , \phi_B=0) 
                   +\sigma(\phi_A=0 , \phi_B=\pi)
\nonumber \\
               &   \! \! \! \! \!
                   -\sigma(\phi_A=0 , \phi_B=\pi/2)
                   -\sigma(\phi_A=0 , \phi_B=3\pi/2)
\nonumber \\
               &   \! \! \! \! \!
                   -\sigma(\phi_A=\pi , \phi_B=0)    
                   -\sigma(\phi_A=\pi , \phi_B=\pi) 
\nonumber \\  
               &   \! \! \! \! \!
                   +\sigma(\phi_A=\pi , \phi_B=\pi/2)    
                   +\sigma(\phi_A=\pi , \phi_B=3\pi/2)  \,  \bigg ] 
\nonumber \\
              & =  \frac{  2 \, sin \theta \, cos \theta \, sin \phi \,
                           U_{2,1}^{T Q_2} }
                 { 2 \, W_{0,0}
                          +  \, (\frac{1}{3}-cos^2 \theta ) \, W_{2,0} }
\ .
\end{align}

We have found that there are a variety of polarization asymmetries in our
Drell-Yan processes in comparison with those in the pp reactions.
We propose that the following asymmetries be measured for finding
the structure functions:
\begin{alignat}{5}
& < \! \sigma \! >, \ \ & & 
A_{LL}, \ \             & &
A_{TT}, \ \             & &
A_{LT}, \ \             & &
A_{TL}, \ \        \nonumber \\
& A_{UT}, \ \           & &
A_{TU},   \ \           & &
A_{UQ_0}, \ \           & &     
A_{TQ_0}, \ \           & &
A_{UQ_1}, \ \      \nonumber \\
& A_{LQ_1}, \ \         & &
A_{TQ_1}, \ \           & &
A_{UQ_2}, \ \           & &
A_{LQ_2}, \ \           & &
A_{TQ_2}.
\end{alignat}
The new structure functions with tensor structure could be found
by the quadrupole polarizations $Q_0$, $Q_1$, and $Q_2$.
For example, the $Q_0$ structure functions could be
found by the asymmetry $A_{UQ_0}$ in Eq. (\ref{eqn:a-uq0}).
It should be associated with the unobserved structure function $b_1$,
so that it is important to measure it in the hadron-hadron reaction.
In addition, there are interesting quadrupole polarizations.
There exist structure functions for the intermediate polarization $Q_1$,
and one of them could be investigated, for example, by the asymmetry 
$A_{UQ_1}$ in Eq. (\ref{eqn:a-uq1}).
The $Q_2$ structure functions are related to the quadrupole polarization
in the transverse plane, and one of them could be measured by 
the asymmetry $A_{UQ_2}$ in Eq. (\ref{eqn:a-uq2}).
In this way, it was revealed that there are many unexplored topics
in the Drell-Yan processes with a spin-1 hadron. 
A parton-model analysis of the structure functions is discussed
in Ref. \cite{our2}. However, further investigations are necessary
on the structure of spin-1 hadrons and relations to various
reactions with them. 

A bonus of our formalism is the possibility of discussing
the $\bar u/\bar d$ asymmetry \cite{skpr} in the longitudinally polarized
and transversity distributions by combining the pp and pd Drell-Yan data.
It is well known that the proton-deuteron Drell-Yan asymmetry is sensitive
to the ratio $\bar u/\bar d$ in the unpolarized case. In fact, the recent
E866 experimental data \cite{e866} clearly showed the ratio $\bar u/\bar d$
as a function of the momentum fraction $x$. In the same way,
it is in principle possible to study the polarized antiquark asymmetries
$\Delta \bar u/\Delta \bar d$ and $\Delta_{_T} \bar u/\Delta_{_T} \bar d$ 
by the pp and pd Drell-Yan processes in addition to the $W$ production
studies (only for $\Delta \bar u/\Delta \bar d$ \cite{js}) at RHIC.
However, the polarized pd formalism had not been available until our
studies. It is now possible to discuss the relation between
the polarized p-d Drell-Yan asymmetry and the polarized flavor asymmetry.

Because the structure functions of the spin-1 hadrons have not been
well studied, it is important to investigate more about their spin
structure theoretically. Furthermore, the experimental possibility of 
the polarized pd reactions should be discussed seriously,  
especially at RHIC \cite{rhic-d}.

%%%%%%%%%%%%%%%%%%%%%%%%%%%%%%%%%%%%%%%%%%%%%%%%%%%%%%%%%%%%%%%%%%%%%%%%%%%%%%%%
%%%%%%%%%%%%%%%%%%%%%%%%%%%%%%%%%%%%%%%%%%%%%%%%%%%%%%%%%%%%%%%%%%%%%%%%%%%%%%%%
\section{Summary}\label{sum}

We have shown that 108 structure functions can be studied in
the polarized Drell-Yan processes with spin-1/2 and spin-1 hadrons.
Because the number is 48 in the reactions of spin-1/2 hadrons,
there are 60 new structure functions.
After integrating over $\vec Q_T$ or after taking the limit
$Q_T\rightarrow 0$, we have shown that 22 structure functions
exist by two independent methods with the requirements of
Hermiticity, parity conservation, and time-reversal invariance.
Because the number is 11 in the processes of spin-1/2 hadrons,
there are 11 new structure functions in the reactions
of spin-1/2 and spin-1 hadrons.
We have shown that these are related to the tensor structure of
the spin-1 hadron and they can be measured by the quadrupole
polarization experiments. A number of spin asymmetries were
shown for extracting the new structure functions. The quadrupole
spin asymmetries with the $Q_0$, $Q_1$, and $Q_2$ polarizations
are valuable for measuring the tensor structure functions.
It is also interesting to find the structure functions for
the intermediate polarizations in the sense that they do not contribute
to the longitudinally and transversely polarized cross sections.
In this paper, we have discussed generally what kind of structure
functions is investigated in the polarized processes with
spin-1/2 and spin-1 hadrons.
We hope that our studies will be materialized experimentally
as the polarized proton-deuteron reactions
in the RHIC-Spin project and also other future projects.

%%%%%%%%%%%%%%%%%%%%%%%%%%%%%%%%%%%%%%%%%%%%%%%%%%%%%%%%%%%%%%%%%%%%%%%%%%%%%%
%%%%%%%%%%%%%%%%%%%%%%%%%%%%%%%%%%%%%%%%%%%%%%%%%%%%%%%%%%%%%%%%%%%%%%%%%%%%%%
\section*{{\bf Acknowledgments}}
\addcontentsline{toc}{section}{\protect\numberline{\S}{Acknowledgments}}

S.K. was partly supported by the Grant-in-Aid for Scientific Research
from the Japanese Ministry of Education, Science, and Culture under
the contract number 10640277. S.H. and S.K. would like to thank
D. E. Soper for communications on the hadron tensor $W^{\mu\nu}$ of
the Drell-Yan process of spin-1/2 hadrons in Refs. \cite{rs,soper}.
They thank E. D. Courant for sending them his report on the possibility
of using polarized deuteron at RHIC and thank K. Kubo for comments on
the polarization of a spin-1 hadron.

%%%%%%%%%%%%%%%%%%%%%%%%%%%%%%%%%%%%%%%%%%%%%%%%%%%%%%%%%%%%%%%%%%%%%%%%%%%%%%
%%%%%%%%%%%%%%%%%%%%%%%%%%%%%%%%%%%%%%%%%%%%%%%%%%%%%%%%%%%%%%%%%%%%%%%%%%%%%%

%%%%%%%%%%%%%%%%%%%%%%%%%%%%%%%%%%%%%%%%%%%%%%%%%%%%%%%%%%%%%%%%%%%%%%%%%%%%%%
%%%%%%%%%%%%%%%%%%%%%%%%%%%%%%%%%%%%%%%%%%%%%%%%%%%%%%%%%%%%%%%%%%%%%%%%%%%%%%

%%%%%%%%%%%%%%%%%%%%%%%%%%%%%%%%%%%%%%%%%%%%%%%%%%%%%%%%%%%%%%%%%%%%%%%%%%%%%%
%%%%%%%%%%%%%%%%%%%%%%%%%%%%%%%%%%%%%%%%%%%%%%%%%%%%%%%%%%%%%%%%%%%%%%%%%%%%%%

\end{document}